\documentclass{article}
\usepackage[latin1]{inputenc}
\usepackage[T1]{fontenc} 
\usepackage{graphicx}
\usepackage{amsmath}
\usepackage{colonequals}
\usepackage{amsthm}
\usepackage{amssymb}
\usepackage{float}
\usepackage{algorithmic}
\usepackage{algorithm}

\newtheorem{definition}{Definition}[section]

\DeclareMathOperator{\Var}{Var}
\DeclareMathOperator{\E}{E}

\DeclareMathOperator{\se}{se}

\DeclareMathOperator{\BIC}{BIC}
\DeclareMathOperator{\Id}{Id}
\DeclareMathOperator{\CI}{CI}

\DeclareMathOperator{\RMSE}{RMSE}
	     
\begin{document}

	\title{\bf Heavy tailed spatial autocorrelation models}
	\date{\small \today}
	\author{A. Kreuzer\footnote{Corresponding author, a.kreuzer@tum.de}, T. Erhardt, T. Nagler,  C. Czado \\
	Zentrum Mathematik, Technische Universit\"at M\"unchen}
	\maketitle

\begin{abstract}
Appropriate models for spatially autocorrelated data account for the fact that observations are not independent. A popular model in this context is the simultaneous autoregressive (SAR) model that allows to model the spatial dependency structure of a response variable and the influence of covariates on this variable. This spatial regression model assumes that the error follows a normal distribution. Since this assumption cannot always be met, it is necessary to extend this model to other error distributions. We propose the extension to the $t$-distribution, the tSAR model, which can be used if we observe heavy tails in the fitted residuals of the SAR model. In addition, we provide a variance estimate that considers the spatial structure of a variable which helps us to specify inputs for our models. An extended simulation study shows that the proposed estimators of the tSAR model are performing well and in an application to fire danger we see that the tSAR model is a notable improvement compared to the SAR model. 
\end{abstract}

\section{Introduction}

\begin{figure}[t]
\center
\includegraphics[width=1\textwidth, trim = 0 2.5cm 0 3cm, clip]{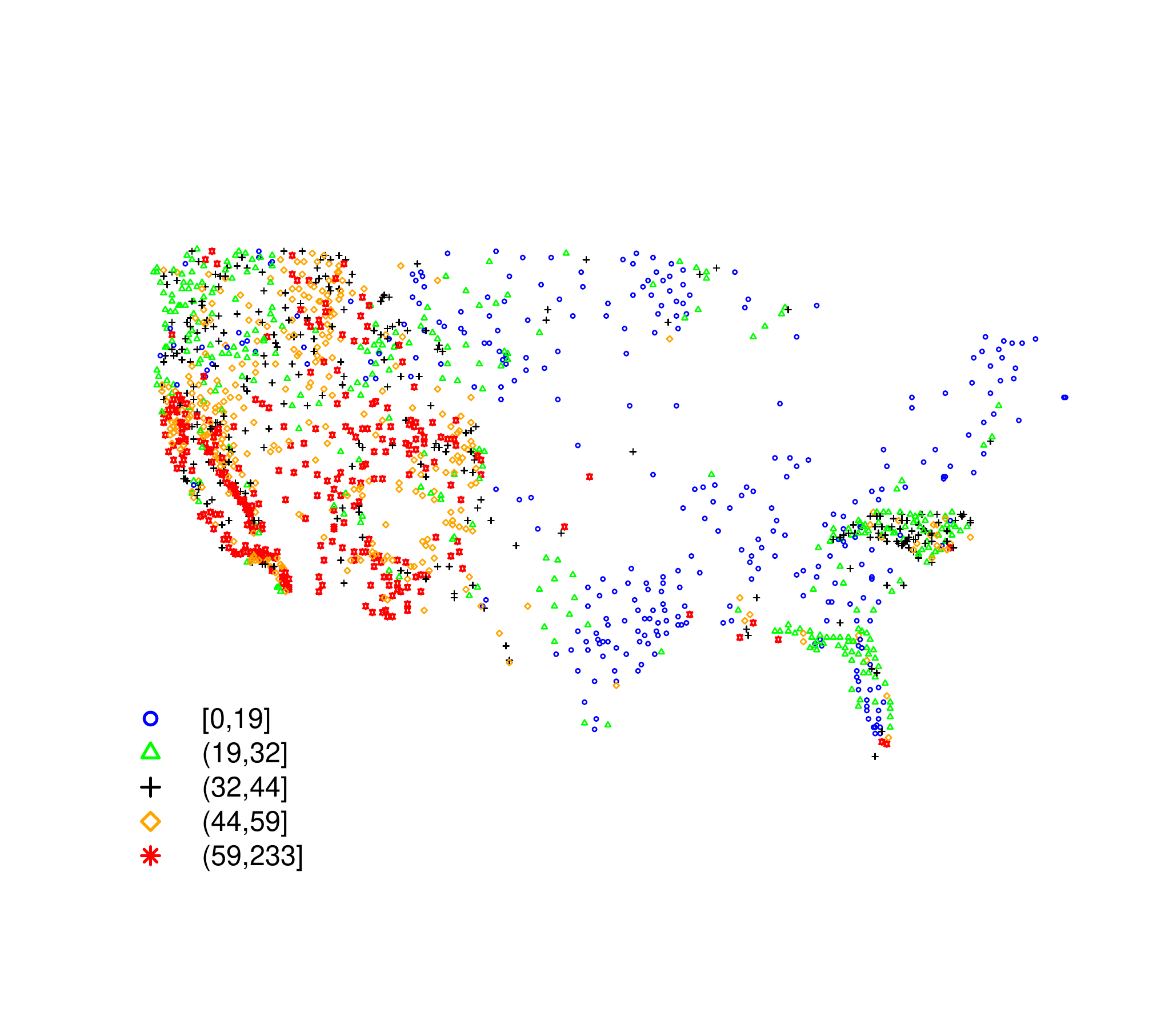}
\caption{Spatial distribution of the Burning Index visualized on the map. We have one observation for every location. The cutpoints of the symbol key are the 20\%, 40\%, 60\% and 80\% quantile of the variable.}
\label{bi}
\end{figure}

``Coincidence of value similarity with locational similarity" is how Anselin and Bera \cite{anselin1998spatial} loosely describe spatial autocorrelation. For illustration we show the Burning Index, a measure for fire danger, for different locations in the US (Figure \ref{bi}). We observe that similar values cluster together, indicating (positive) spatial autocorrelation. Spatial autocorrelation occurs in many different types of data, for example in climate (fire danger, droughts) or economics (unemployment) data. This is why statistical methods that can deal with spatial autocorrelation are of high interest. A first contribution to this field was made by Whittle \cite{whittle1954stationary} who provided a framework for stochastic processes on the plane. Whittle introduced autoregressive models in two dimensions. Following this idea, Ord \cite{ord1975estimation} proposed the simultaneous autoregressive (SAR) model. This model not only allows us to capture the spatial dependency structure of a response variable but also the influence of covariates on this variable. This property of the SAR model makes it very attractive and led to extensions. Pace and Barry \cite{pace1997sparse} studied how sparse spatial weight matrices can speed up the estimation procedure and De Olivera and Song \cite{de2008bayesian} provide a Bayesian framework for the SAR model.

This work was motivated by an attempt to investigate the influence of weather conditions on fire danger in the continental US while accounting for spatial dependency. Data are obtained from the Wildland Fire Assessment System (WFAS). WFAS generates maps for observed and forecasted weather, fuel moisture and fire danger in the US.
The SAR model is based on the assumption that the error follows a normal distribution, an assumption that cannot always be met. In our fitted model we observed residuals having heavier tails than the normal distribution. This is why we propose an extension of the SAR model to allow for a $t$-distributed error. We call this the tSAR model (Section \ref{tsardef}). We show how parameters of the tSAR model can be estimated and how the fitted model can be used for prediction (Sections \ref{paresttsar} and \ref{pr}). Furthermore, we provide a spatially varying variance estimate which serves as input to our models (Section \ref{sigmaeps}). In a simulation study (Section \ref{sim}), we show that our proposed estimators for the tSAR model are reasonable and the application (Section \ref{application}) shows that the model fit can improve on the standard SAR model.

\section{The SAR model}
\label{sarmodel}

We recall some basic concepts related to the SAR model. First, we need to be able to determine how certain locations are related to each other, i.e., if there is a link between them and, if so, how strong the connection is. This is usually encoded in a proximity matrix (cf., Waller and Gotway \cite{waller2004applied} p.224 ff.). 
For \(n\) spatial locations \(l_1, ...., l_n \), the \emph{proximity matrix} is a \(n \times n\) matrix where entry \((i,j)\) indicates if and how strong location \(l_i\) is connected to location \(l_j\). A value of zero means that there is no connection from \(l_i\) to \(l_j\). The diagonal of the proximity matrix is set to zero such that a location is not connected to itself. Since this matrix does not need to be symmetric, we need to distinguish between a connection from \(l_i\) to \(l_j\) and a connection from \(l_j\) to \(l_i\).

For a given proximity matrix \(W\) with entries $w_{ij}$, we can introduce the neighbors of location \(l_i\) which are all locations \(l_j\) such that \(w_{ij} \neq 0\). We denote the \emph{set of neighbors of location \(i\)} by \(N_i\), i.e.,
\begin{equation*}
N_i \colonequals \{j \in \{1, \ldots n\}|w_{ij} \neq 0\}.
\end{equation*}

We now provide two possible choices of proximity matrices. In both cases we measure the strength of a connection by the inverse distance between the two corresponding locations. We will use the great circle distance (cf., Banerjee \cite{banerjee2005geodetic}) since our locations are specified as longitude/latitude pairs. For the first example we consider \(N_i(k)\) the set of the \(k\) nearest neighbors of \(l_i\), i.e., the \(k\) locations (excluding \(l_i\)) which have the smallest distance to \(l_i\). Let \(d_{ij}\) denote the distance between \(l_i\) and \(l_j\). 
For given \(k\), entry \((i,j)\) of the \emph{non-standardized nearest neighbors based proximity matrix \(\tilde W\)} is then given by
\begin{equation*}
  \tilde w_{ij} \colonequals \left\{\begin{array}{cl} \frac{1}{d_{ij}}, & \mbox{if } l_j \in N_i(k)     \\ 0, & \mbox{else} \end{array}\right.  ,
\end{equation*}
and entry \((i,j)\) of the \emph{row-standardized nearest neighbors based proximity matrix \(W\)} is defined by
\begin{equation*}
w_{ij} \colonequals \frac{\tilde w_{ij}}{\tilde w_{i.}},
\end{equation*}
where \(\tilde w_{i.} = \sum_{j=1}^n \tilde w_{ij}\) is the sum of the \(i\)-th row of the non-standardized nearest neighbors based proximity matrix \(\tilde W\).
By defining the proximity matrix in this way, we ensure that each location has the same number of neighbors. This is no longer the case if we use a radius to determine the set of neighbors.

For a given radius \(r\), entry \((i,j)\) of the \emph{non-standardized radius based proximity matrix \(\tilde W\)} is given by
\begin{equation*}
\tilde w_{ij} \colonequals \left\{\begin{array}{cl} \frac{1}{d_{ij}}, & \mbox{if } i \neq j ~ \mbox{and} ~  d_{ij} \le r     \\ 0, & \mbox{else} \end{array}\right. .
\end{equation*}
 As above, entry \((i,j)\) of the \emph{row-standardized radius based proximity matrix \(W\)} is then defined by
\begin{equation*}
w_{ij} \colonequals \frac{\tilde w_{ij}}{\tilde w_{i.}},
\end{equation*}
where \(\tilde w_{i.} = \sum_{j=1}^n \tilde w_{ij}\).
A property of radius based proximity matrices is that they are symmetric which is not necessarily the case for nearest neighbors based proximity matrices.
In the following, we will always consider row-standardized proximity matrices and refer to them as \emph{nearest neighbors matrices} and \emph{radius matrices}. This standardization allows us to consider a sum of values weighted with the corresponding entry of the proximity matrix as a weighted average, as we will see in the SAR model.

 In the following we recall the classical SAR model. By $\boldsymbol Z \sim N_n(\boldsymbol \mu, S)$ we denote that the random vector $\boldsymbol Z$ follows a $n$-dimensional normal distribution with mean vector $\boldsymbol \mu$ and covariance matrix $S$.

\begin{definition}[The simultaneous autoregressive (SAR) model]
Let \(\boldsymbol Y = (Y_1,...,Y_n)^T\) be a \(n\)-dimensional random vector and \(x_{i1},...,x_{ip}\) for \(i=1, ..., n\) associated (fixed) covariates. Let \(X \in \mathbb{R}^{n \times (p+1)}\) be a matrix whose \(i\)-th row is given by \(\boldsymbol x_i^T\), \( \boldsymbol x_i \colonequals (1,x_{i1},...,x_{ip})^T  \). Then the \emph{simultaneous autoregressive (SAR) model} is given by
\begin{equation}
\boldsymbol Y = X \boldsymbol\beta + \lambda W(\boldsymbol Y - X \boldsymbol\beta) + \boldsymbol\epsilon ,
\label{sar}
\end{equation}
where \( \lambda \in \mathbb{R}\) is the spatial dependence parameter, \( W \in \mathbb{R}^{n \times n} \) is the proximity matrix and \( \boldsymbol\beta \in \mathbb{R}^{p+1} \) the unknown regression coefficient. For the error vector we assume \(\boldsymbol\epsilon \sim N_n(0,\sigma^2\Sigma_{\epsilon})\) with a positive scalar \(\sigma\) and a diagonal matrix \( \Sigma_{\epsilon} \in \mathbb{R}^{n \times n}\) with positive diagonal entries.
\label{sardef}
\end{definition}
 So the components of \(\boldsymbol\epsilon\) are independent. In our application we need to allow for different error variances per location, i.e., the diagonal elements of $\Sigma_{\epsilon}$ are different. Furthermore we require the matrix \( (\Id_n - \lambda W) \) to be a full rank matrix in order to ensure that the model is well defined. Here $\Id_n$ denotes the $n$-dimensional identity matrix.

Writing Equation \eqref{sar} component wise yields
\begin{equation}
\begin{split}
Y_i &=  \boldsymbol \beta^T \boldsymbol x_i + \lambda \sum_{j \in N_i} w_{ij} (Y_j - \boldsymbol \beta^T \boldsymbol x_j ) + \epsilon_i \text{ for } i = 1, ..., n , 
\end{split}
\label{sarcomp}
\end{equation}
where \(N_i = \{j|w_{ij} \neq 0   \}\) is the set of neighbors of the \(i\)-th location as introduced above. As we consider row-standardized proximity matrices, the \emph{spatial component} $\lambda \sum_{j \in N_i} w_{ij} (Y_j - \boldsymbol \beta^T \boldsymbol x_j ) $ can be seen as a weighted average of the deviations of the \emph{linear component} $X \boldsymbol \beta $ from the response in the corresponding neighborhood.
In the following, we always assume the proximity matrix $W$ and $\Sigma_{\epsilon}$ to be known.

\subsection[Parameter estimation]{Parameter estimation}
We briefly sketch how parameters of the SAR model are estimated since we want to approach parameter estimation for the tSAR model in similar way. We follow Waller and Gotway \cite{waller2004applied} (p. 365 ff.) who estimate the parameters by maximizing the likelihood. This requires to derive the likelihood function.

Since \( (\Id_n - \lambda W) \) has full rank, we can express Equation \eqref{sar} as
\begin{equation}
\boldsymbol Y = (\Id_n - \lambda W)^{-1} \boldsymbol\epsilon + X \boldsymbol\beta, 
\label{ssaralty}
\end{equation}
and we see that \(\boldsymbol Y\) (as a full rank linear transformation of a normal random variable) is normally distributed with mean vector
\begin{equation*}
\E(\boldsymbol Y) = X \boldsymbol\beta ,
\end{equation*}
and covariance matrix
\begin{equation}
\Var(\boldsymbol Y) =  \sigma^2 \Sigma_Y(\lambda),
\label{ssarvar}
\end{equation}
where \(\Sigma_Y(\lambda) := (\Id_n - \lambda W)^{-1} \Sigma_{\epsilon} (\Id_n - \lambda W^T)^{-1}\). 

Knowing the distribution of \(\boldsymbol Y\), the likelihood function for $(\boldsymbol\beta,\sigma,\lambda)$ for given data $\boldsymbol y$ is given by
\begin{equation*}
\begin{split}
L(\boldsymbol y|\boldsymbol\beta,\sigma,\lambda) \colonequals & (2\pi)^{-\frac{n}{2}} \det[\sigma^2 \Sigma_{Y}(\lambda)]^{-\frac{1}{2}} \cdot \\
& \cdot \exp\left[-\frac{1}{2}(\boldsymbol y - X \boldsymbol\beta)^T \frac{1}{\sigma^2} \Sigma_{Y}(\lambda)^{-1} (\boldsymbol y - X\boldsymbol\beta)   \right].
\end{split}
\end{equation*}
Instead of maximizing the likelihood function, we minimize the negative log-likelihood given by
\begin{equation}
\begin{split}
\ell(\boldsymbol y|\boldsymbol\beta,\sigma,\lambda) \colonequals & - \log\left[L(\boldsymbol y|\boldsymbol\beta,\sigma,\lambda)\right] \\
=& \frac{n}{2} \log(2\pi) + \frac{n}{2}\log(\sigma^2) + \frac{1}{2}\log\left\{\det[\Sigma_Y(\lambda)]\right\} + \\
&+ \frac{1}{2\sigma^2}   (\boldsymbol y-X\boldsymbol\beta)^T \Sigma_Y(\lambda)^{-1} (\boldsymbol y-X\boldsymbol \beta) .
\end{split}
\label{ssarnll}
\end{equation}

\subsubsection*{Estimation of $\boldsymbol\beta$}
First we take the derivative of \(\ell(\boldsymbol y|\boldsymbol\beta,\sigma,\lambda)\) with respect to \(\boldsymbol\beta\) and set it to zero. Solving for \(\boldsymbol\beta\) yields the (on \(\lambda\) dependent) estimate
\begin{equation}
\hat{\boldsymbol\beta}(\lambda) = \left[X^T \Sigma_Y(\lambda)^{-1} X \right]^{-1}     X^T \Sigma_Y(\lambda)^{-1} \boldsymbol y ,
\label{ssarbeta}
\end{equation}
which is independent of \(\sigma\). For fixed $\lambda$, this is the generalized least squares estimator for $\boldsymbol\beta$ (cf., Kariya and Kurata\cite{kariya2004generalized} p. 35). 
\subsubsection*{Estimation of $\sigma$}
We proceed in the same way for \(\sigma^2\) and obtain the (on \(\boldsymbol\beta\) and \(\lambda\) dependent) estimate
\begin{equation}
\hat\sigma^2(\boldsymbol\beta, \lambda) =\frac{1}{n}(\boldsymbol y-X\boldsymbol\beta)^T \Sigma_Y(\lambda)^{-1} (\boldsymbol y-X\boldsymbol \beta).
\label{ssarsigmaest}
\end{equation}
The estimate for \(\sigma\) is given by its positive square root, i.e.,
\begin{equation*}
\hat\sigma(\boldsymbol\beta, \lambda) =\sqrt{\frac{1}{n}(\boldsymbol y-X\boldsymbol\beta)^T \Sigma_Y(\lambda)^{-1} (\boldsymbol y-X\boldsymbol \beta)}.
\end{equation*}

\subsubsection*{Estimation of $\lambda$}
There is no closed form solution for \( \lambda  \). So we focus on the negative profile log-likelihood given by
\begin{equation*}
\begin{split}
&\frac{n}{2} \log(2\pi) + \frac{n}{2}\log \left[ \hat{\sigma}(\hat{\boldsymbol\beta}(\lambda), \lambda)^2  \right] + \frac{1}{2}\log\left\{\det[\Sigma_Y(\lambda)]\right\}  + \\
&+ \frac{\left[y-X\hat{\boldsymbol\beta}(\lambda)\right]^T \Sigma_Y(\lambda) \left[y-X\hat{\boldsymbol\beta}(\lambda)\right]}{2\hat{\sigma}(\hat{\boldsymbol\beta}(\lambda), \lambda)^2} ,
\end{split}
\end{equation*}
which is obtained by replacing \(\boldsymbol\beta\) by \(\hat {\boldsymbol{\beta}}(\lambda) \) and \(\sigma\) by \( \hat{\sigma}(\hat{\boldsymbol\beta}(\lambda), \lambda) \) in the negative log-likelihood function (\ref{ssarnll}). This one dimensional nonlinear minimization problem can be solved by appropriate optimization algorithms and yields \(\hat\lambda\), the estimate of \(\lambda\). The estimation procedure is implemented in the \(\texttt{R}\) package \(\texttt{spdep}\) (see Bivand \cite{bivandspdep}). For optimization, the \(\texttt{R}\) function \(\texttt{optimize}\) which is a combination of golden section search and successive parabolic interpolation (see Brent \cite{brent1973algorithms}) is used.
The final estimate of \(\boldsymbol\beta\) is then given by \(\hat{\boldsymbol\beta} = {\boldsymbol\beta}(\hat\lambda)  \) and the final estimate of \(\sigma\) is given by \(\hat\sigma = \hat\sigma(\hat{\boldsymbol\beta}, \hat\lambda)\).

\subsection{Prediction and residuals}
From Equation \eqref{sarcomp} it follows that the conditional expectation of $\boldsymbol Y$ at spatial location $i$, given the values of all other spatial locations, is 
\begin{equation*}
\begin{split}
\E(Y_i|\boldsymbol Y_{-i} = \boldsymbol y_{-i}) &= \E(Y_i|Y_j = y_j, j \in N_i) \\
&= \boldsymbol \beta^T \boldsymbol x_i + \lambda \sum_{j \in N_i} w_{ij} (y_j - \boldsymbol \beta^T \boldsymbol x_j ) ,
\end{split}
\end{equation*}
where $\boldsymbol z_{-i} = \{z_1, \ldots z_n\} \setminus \{z_i\}$ for a $n-$dimensional vector $\boldsymbol z$.  
So we define the \emph{\(i\)-th local prediction of Y}, where the neighbors' values are observed, by
\begin{equation*}
\hat y_{i|N_i} \colonequals \hat{\boldsymbol \beta}^T \boldsymbol x_i + \hat\lambda \sum_{j \in N_i} w_{ij} (y_j - \hat{\boldsymbol \beta}^T \boldsymbol x_j ) ,
\end{equation*}
and the corresponding \emph{vector of local predictions} is defined by
\begin{equation*}
\hat {\boldsymbol y}_{|N} \colonequals X \hat{\boldsymbol\beta} + \hat\lambda W(\boldsymbol y - X \hat{\boldsymbol\beta}) . 
\end{equation*}

Based on the prediction we can define the \emph{\(i\)-th local residual} as
\begin{equation*}
 \hat \epsilon_i \colonequals y_i - \hat y_{i|N_i}.
\end{equation*}
Since the local residual is the only type of residual we consider, we also refer to it just as the \emph{\(i\)-th residual}. 
The \emph{\(i\)-th standardized residual} is given by
\begin{equation*}
\tilde{\epsilon}_i \colonequals \frac{\hat \epsilon_i}{\sqrt{\hat\sigma^2 (\Sigma_{\epsilon})_{ii}}},
\end{equation*}
since \(\Var(\epsilon_i) = \sigma^2 (\Sigma_{\epsilon})_{ii}\).
From a good fit we expect the standardized residuals to be approximately identically and independent standard normally distributed.

Furthermore, an estimate for the \emph{standard error of \(\hat\beta_i\)} is provided by
\begin{equation*}
\hat\se(\hat\beta_i) \colonequals \hat\sigma \sqrt{ \left(\left[X^T \Sigma_Y(\hat\lambda)^{-1} X\right]^{-1}\right)_{ii}},
\end{equation*}
since
 \begin{equation}
 \begin{split}
 \Var(\hat{\boldsymbol \beta}(\lambda)) =& \Var\left( \left[X^T \Sigma_Y(\lambda)^{-1} X \right]^{-1}     X^T \Sigma_Y(\lambda)^{-1} \boldsymbol Y\right) \\
 =& \left[X^T \Sigma_Y(\lambda)^{-1} X \right]^{-1}     X^T \Sigma_Y(\lambda)^{-1} \Var(\boldsymbol Y) \cdot \\
 & \cdot \left\{\left[X^T \Sigma_Y(\lambda)^{-1} X \right]^{-1}     X^T \Sigma_Y(\lambda)^{-1}\right\}^T \\
 =& \sigma^2 \left[X^T \Sigma_Y(\lambda)^{-1} X\right]^{-1} .
 \end{split}
 \label{ssarvarbeta}
 \end{equation}

This can be used to test the significance of \(\beta_i\).
For fixed \(\lambda\), \( \boldsymbol{ \hat\beta}\) is normally distributed (as a linear transformation of the normally distributed vector \(\boldsymbol Y\)).
We use the following test for the significance of \(\beta_i\) with significance level \(\alpha\), null hypothesis \(H_0: \beta_i = 0\) and alternative \(H_1: \beta_i \neq 0\).
We reject \(H_0\) if 
\begin{equation*}
\left |\frac{\hat\beta_i}{\hat\se(\hat\beta_i)}\right | > \Phi^{-1}(1-\frac{\alpha}{2}) ,
\end{equation*}
where \(\Phi^{-1}(1-\frac{\alpha}{2})\) denotes the \(1-\frac{\alpha}{2}\) quantile of the $N(0,1)$ distribution.
But we need to use this test with caution because the standard error was estimated with the assumption that $\lambda$ was known. Thus the standard error is too small since we do not account for the variation in \(\lambda\).

\section{The tSAR model}
\label{tsar}

The tSAR model is a way of extending the SAR model to allow for a Student \(t\) error distribution. We replace the assumption that the error vector is normally distributed by the assumption that the components of the error vector are univariate $t$-distributed. This allows for heavier tailed errors in our model.  
\subsection{Model definition}
\label{tsardef}
We say that the one dimensional random variable \(X\) follows a \emph{\(t\)-distribution} with mean \(\mu ~ (\mu \in \mathbb{R})\), scale parameter \(\Sigma ~ (\Sigma  \in \mathbb{R}, \Sigma >0)\) and \(\nu\) (\(\nu \in \mathbb{N}\)) degrees of freedom if \(X\) has the density
\begin{equation*}
t( x|\mu, \Sigma, \nu) \colonequals \Gamma\left(\frac{\nu + 1}{2}\right) \Gamma\left(\frac{\nu}{2}\right)^{-1}  \frac{1}{ \sqrt{ \nu \pi \Sigma }}\left[1 + \frac{( x - \mu)^2}{\nu \Sigma}     \right]^{-\frac{\nu + 1}{2}},
\end{equation*}
where \(\Gamma(x) \colonequals \int_0^{\infty}s^{x-1}e^{-s}ds\) is the gamma function. We write \( X \sim t( \mu, \Sigma, \nu) \). Furthermore, we denote by $Sc( X)$ the scale parameter of $X$.
\label{tdist}
According to Kotz and Nadarajah\cite{kotz2004multivariate} (p. 10 ff.) it holds that
\begin{equation}
\E(X) = \mu ,
\end{equation}
and
\begin{equation}
\Var(X) = \frac{\nu}{\nu -2} \Sigma ,
\end{equation}
for $\nu > 2$.
\begin{definition}[tSAR model]
In the \emph{tSAR model} we assume that
\begin{equation*}
\boldsymbol Y=  X \boldsymbol\beta + \lambda  W(Y - X \boldsymbol\beta) + \boldsymbol\epsilon ,
\end{equation*}
with  \(\epsilon_i \sim t(0,\sigma^2 (\Sigma_{\epsilon})_{ii}, \nu)\) with a positive scalar \(\sigma\) and a diagonal matrix \( \Sigma_{\epsilon} \in \mathbb{R}^{n \times n}\) with positive diagonal entries and \(\nu > 2  \) degrees of freedom. Furthermore, we assume that the components of the vector \(\boldsymbol\epsilon\) are independent. \(X, \lambda, W\) and \(\boldsymbol \beta\) are defined as in Definition \ref{sardef}.
\end{definition}

\subsection[Parameter estimation]{Parameter estimation}
\label{paresttsar}
As in the SAR model, we estimate parameters by maximizing the likelihood while assuming $W, \Sigma_{\epsilon}$ and the degrees of freedom $\nu$ to be known.  We start with deriving the likelihood function.
Since
\begin{equation*}
\boldsymbol\epsilon = (\Id_n - \lambda W) (\boldsymbol Y - X \boldsymbol\beta) ,
\end{equation*}
the components of the vector
\begin{equation*}
\boldsymbol Z := \Sigma_{\epsilon}^{-\frac{1}{2}}(\Id_n - \lambda W) (\boldsymbol Y - X \boldsymbol\beta),
\end{equation*} 
where \(\Sigma_{\epsilon}^{-\frac{1}{2}}\) is a diagonal matrix with \(i\)-th diagonal entry \((\Sigma_{\epsilon})_{ii}^{-\frac{1}{2}}\), are identically and independent \(t(0,\sigma^2, \nu)\) distributed.
So the density \(f_{\boldsymbol Z}\) of \( \boldsymbol Z\) is the product of its marginal densities.
Furthermore, we have that
\begin{equation*}
\boldsymbol Y =  (\Id_n - \lambda W)^{-1} \sqrt{\Sigma_{\epsilon}} \boldsymbol Z + X \boldsymbol\beta .
\end{equation*}
We obtain the density of \( \boldsymbol Y \) by density transformation.
\begin{equation*}
\begin{split}
f_Y (\boldsymbol y)=& \left|{\det\left[\Sigma_{\epsilon}^{-\frac{1}{2}}(\Id_n - \lambda W) \right]}\right| \prod_{i=1}^n t\left( \left(\Sigma_{\epsilon}^{-\frac{1}{2}}(\Id_n - \lambda W) (\boldsymbol y - X \boldsymbol\beta)\right)_i| 0, \sigma^2, \nu  \right) ,
\end{split}
\end{equation*}
where \(f_{\boldsymbol Z}\) is the density of \(\boldsymbol Z\). 
Hence the negative log-likelihood of data $\boldsymbol y$ given the model parameters $(\boldsymbol\beta,\sigma,\lambda)$ is 

\begin{equation}
\begin{split}
\ell(\boldsymbol y|\boldsymbol\beta, \lambda, \sigma) \colonequals& -\log\left\{\left|{\det\left[\Sigma_{\epsilon}^{-\frac{1}{2}}(\Id_n - \lambda W) \right]}\right|\right\}\\
&- \sum_{i=1}^n \log\left[t\left( \left(\Sigma_{\epsilon}^{-\frac{1}{2}}(\Id_n - \lambda W) (\boldsymbol y - X \boldsymbol\beta)\right)_i| 0, \sigma^2, \nu  \right) \right].
\label{tsarnll}
\end{split}
\end{equation}
\newline
Unfortunately we can not proceed as before (i.e., take the derivatives with respect to \(\boldsymbol\beta  \) and \(\sigma\), set them to zero and solve analytically for the parameters) due to the more complex form of the likelihood function. For illustration of this problem we write down the derivative with respect to \(\boldsymbol \beta\).
\begin{equation*}
\frac{d}{d\boldsymbol\beta} \ell(\boldsymbol y|\boldsymbol\beta, \lambda, \sigma) = \text{c} + \frac{\nu+1}{2} \sum_{i=1}^n  \frac{1}{\left[1 + \frac{m_i(\boldsymbol\beta)^2}{\sigma^2\nu}   \right]} \frac{2 m_i(\boldsymbol\beta)}{\sigma^2\nu}  \left( \Sigma_{\epsilon}^{-\frac{1}{2}}(\Id_n - \lambda W) X \right)_i ,
\end{equation*}
where \(\text{c}\) is a constant independent of \(\boldsymbol \beta\), \(m_i(\boldsymbol\beta) = \left(\Sigma_{\epsilon}^{-\frac{1}{2}} (\Id_n - \lambda W) (\boldsymbol y - X \boldsymbol\beta)\right)_i \) and \(\left( \Sigma_{\epsilon}^{-\frac{1}{2}}(\Id_n - \lambda W) X \right)_i\) is the \(i\)-th row of \(\Sigma_{\epsilon}^{-\frac{1}{2}}(\Id_n - \lambda W) X\). If we set this equation to zero, we can not solve it analytically for \(\boldsymbol \beta\). Numerical optimization for all parameters would be computationally very complex since $\boldsymbol\beta$ is often high dimensional. Therefore we suggest to estimate \(\boldsymbol\beta\) and \(\sigma\) as explained in the following.

\subsubsection*{Estimation of $\boldsymbol \beta$}
A simple analytic estimator for \(\boldsymbol \beta\) is the on \(\lambda\) dependent generalized least squares estimator, i.e.,
\begin{equation*}
\hat{\boldsymbol\beta}(\lambda) = \left[X^T \Sigma_Y(\lambda)^{-1} X \right]^{-1}     X^T \Sigma_Y(\lambda)^{-1} \boldsymbol Y ,
\end{equation*}
as in the SAR model. For fixed $\lambda$, this is the best linear unbiased estimator according to the Gau{\ss} Markov Theorem (cf., Kariya and Kurata \cite{kariya2004generalized} p. 34).     

\subsubsection*{Estimation of $\sigma$}
For \(\sigma\) we suggest the following estimate dependent on \(\boldsymbol\beta\) and \(\lambda\),
\begin{equation*}
\hat\sigma^2(\boldsymbol\beta, \lambda) = \frac{\nu-2}{\nu}   \frac{1}{n}(\boldsymbol y-X\boldsymbol\beta)^T \Sigma_Y(\lambda)^{-1} (\boldsymbol y-X\boldsymbol \beta),
\end{equation*}
since \(\hat\sigma^2(\boldsymbol\beta, \lambda)\)  can be written as
\begin{equation*}
\begin{split}
\hat\sigma^2(\hat{\boldsymbol\beta}, \hat\lambda) &= \frac{\nu-2}{\nu}  \frac{1}{n}(\boldsymbol y-X\hat{\boldsymbol\beta})^T \Sigma_Y(\hat\lambda)^{-1} (\boldsymbol y-X\hat{\boldsymbol\beta}) \\
&=  \frac{\nu-2}{\nu} \frac{1}{n}(\boldsymbol y-X\hat{\boldsymbol\beta})^T (\Id_n - \hat\lambda W^T) \Sigma_{\epsilon}^{-1} (\Id_n - \hat\lambda W) (\boldsymbol y-X\hat{\boldsymbol\beta}) \\
&=   \frac{\nu-2}{\nu} \frac{1}{n} \hat{\boldsymbol\epsilon}^T \Sigma_{\epsilon}^{-1} \hat{\boldsymbol\epsilon} \\
&=  \frac{\nu-2}{\nu} \frac{1}{n}\sum_{i=1}^n \frac{\hat\epsilon_i^2}{(\Sigma_{\epsilon})_{ii}} ,
\end{split}
\end{equation*}
where we used the definition of the prediction vector $ \hat {\boldsymbol y}_{|N}  \colonequals X \hat{\boldsymbol\beta} + \hat\lambda W(\boldsymbol y - X \hat{\boldsymbol\beta})$ and the residual $\hat{ \boldsymbol \epsilon} \colonequals \boldsymbol y - \hat{ \boldsymbol y}_{|N}$
to express $\hat{  \boldsymbol\epsilon}$ as
\begin{equation*}
\begin{split}
\hat{  \boldsymbol\epsilon} &= \boldsymbol y - \hat { \boldsymbol y}_{|N} \\
&= \boldsymbol y -  X \hat{\boldsymbol\beta} - \hat\lambda W \boldsymbol y + \hat\lambda W X \hat{\boldsymbol\beta} \\
&= (\Id_n - \hat\lambda W)(\boldsymbol y -  X \hat{\boldsymbol\beta}).
\end{split}
\end{equation*}

The quantity \(\frac{1}{n}\sum_{i=1}^n \frac{\hat\epsilon_i^2}{(\Sigma_{\epsilon})_{ii}} \) is an estimate of the variance of  \( \epsilon_i/\sqrt{(\Sigma_{\epsilon})_{ii}}\) and so \(\hat\sigma^2 (\hat{\boldsymbol\beta}, \hat\lambda)\) is an estimate of the scale parameter.

\subsubsection*{Estimation of $\lambda$}
For the estimation of \(\lambda\) we proceed as in the SAR model, i.e., we obtain the negative profile log-likelihood by replacing \(\boldsymbol\beta\) by \( \hat{\boldsymbol \beta}(\lambda) \) and \(\sigma\) by \(\hat\sigma(\hat{\boldsymbol\beta}(\lambda), \lambda)\) in the negative log-likelihood function (\ref{tsarnll}). Then \(\hat\lambda\) is defined as the minimum of the negative profile log-likelihood which is found numerically. As before we set \(\hat{\boldsymbol\beta} = \hat{\boldsymbol\beta}(\hat\lambda)  \) and \(\hat\sigma = \hat\sigma(\hat{\boldsymbol\beta}, \hat\lambda)\).

\subsection{Prediction and residuals}
\label{pr}
The \emph{vector of local predictions \(\hat{\boldsymbol y}_{|N}\)} and the \emph{residual vector \(\hat{\boldsymbol \epsilon}\)} are defined as for the SAR model, i.e.,
\begin{equation*}
\hat {\boldsymbol y}_{|N} \colonequals X \hat{\boldsymbol\beta} + \hat\lambda W(\boldsymbol y - X \hat{\boldsymbol\beta}) ,
\end{equation*}
and
\begin{equation*}
 \hat \epsilon_i \colonequals y_i - \hat y_{i|N_i}.
\end{equation*}

Since \(Sc(\epsilon_i) = \sigma^2 (\Sigma_{\epsilon})_{ii}    \), we define the \emph{\(i\)-th standardized residual} by
\begin{equation*}
\tilde{\epsilon}_i \colonequals \frac{\hat \epsilon_i}{\sqrt{\hat\sigma^2 (\Sigma_{\epsilon})_{ii}}}.
\end{equation*}

As in (\ref{ssaralty}), we can write \(\boldsymbol Y\) as
\begin{equation*}
\boldsymbol Y = (\Id_n - \lambda W)^{-1} \boldsymbol\epsilon + X \boldsymbol\beta ,
\end{equation*}
and obtain similarly to Equation (\ref{ssarvar}),
\begin{equation*}
\Var(\boldsymbol Y) = \frac{\nu}{\nu - 2}\sigma^2 \Sigma_Y(\lambda),
\end{equation*}
where \(\Sigma_Y(\lambda) := (\Id_n - \lambda W)^{-1} \Sigma_{\epsilon} (\Id_n - \lambda W^T)^{-1}\). Therefore we get similar to  (\ref{ssarvarbeta})
 \begin{equation*}
 \Var(\hat{ \boldsymbol\beta}(\lambda)) = \frac{\nu}{\nu - 2}\sigma^2 (X^T \Sigma_Y(\lambda)^{-1} X)^{-1} ,
 \end{equation*}
and thus we estimate the \emph{standard error of \(\hat\beta_i\)} by
\begin{equation*}
\hat\se(\hat\beta_i) \colonequals \sqrt{\frac{\nu}{\nu - 2}\hat\sigma^2  ((X^T \Sigma_Y(\hat\lambda)^{-1} X)^{-1})_{{ii}}}.
\end{equation*}

\subsection{Specifying the matrix $\Sigma_{\epsilon}$}
\label{sigmaeps}
To estimate a SAR or tSAR model we need to specify the matrix $\Sigma_{\epsilon}$ which is proportional to the covariance matrix of the error vector. One possibility would be to choose this equal to the identity matrix which leads to all locations having the same variance. But we also want to account for different error variances. Therefore we provide a variance estimate which uses the restriction to a neighborhood. We define the \emph{local empirical variance} of the spatial variable \(Z_i\) at location \(i\) with respect to the proximity matrix \(W\) as follows
\begin{equation}
\hat\sigma^2_{W}(Z_i) \colonequals \frac{1}{|N_i| - 1} \sum_{j \in N_i} (z_j - \bar z_{N_i})^2,
\label{empvar}
\end{equation}
where the \(z_j\) are observations of \(Z\), \(N_i = \{j|w_{ij} \neq 0\} \)  is the neighborhood of location \(i\) induced by \(W\), \(|N_i|\) is the cardinality of the set \(N_i\) and \( \bar z_{N_i} = \frac{1}{|N_i|}\sum_{j \in N_i} z_j\). The corresponding \emph{local empirical variance matrix} is a diagonal matrix with \(i\)-th diagonal entry equal to \(\hat\sigma^2_{W}(Z_i)\).
\label{empvardef}
For a SAR or tSAR model with response $Y$, covariates $x_1,\ldots, x_p$ and proximity matrix W, we propose to specify $\Sigma_{\epsilon}$ in the following way. 
\begin{enumerate}
\item We fit a linear regression model with response variable $Y$ and covariates $x_1,\ldots, x_p$, i.e., we assume
\begin{equation*}
y_i = (x_{1i},\ldots, x_{pi})\boldsymbol \beta_{lm} + \epsilon_{lm,i} 
\end{equation*}
with $\epsilon_{lm,i} \sim N(0,\sigma^2)$, $\boldsymbol \beta_{lm} \in \mathbb{R}^p, \sigma \in \mathbb{R}_+$. We obtain   $\hat{\boldsymbol\beta}_{lm}$, the estimate of $\boldsymbol\beta_{lm}$ by least squares estimation. The $i$-th residual $r_i$ is given by 
\begin{equation*}
r_i = y_i - (x_{1i},\ldots, x_{pi})\hat{\boldsymbol\beta}_{lm}.
\end{equation*}
\item Then we set $\Sigma_{\epsilon}$ equal to the local empirical variance matrix of the residual vector $\boldsymbol r$ with respect to $W$. So $\Sigma_{\epsilon}$ is a diagonal matrix with $i-$th diagonal entry $(\Sigma_{\epsilon})_{ii} = \hat\sigma^2_W(r_i)$, where $\hat\sigma^2_W(\cdot)$ is generally defined in \eqref{empvar}. We call this the \emph{local regression variance matrix of $Y$ with respect to $W$}. 
\end{enumerate}

\section{Simulation study}
\label{sim}

\begin{figure}[H]
\center
\includegraphics[width=1\textwidth]{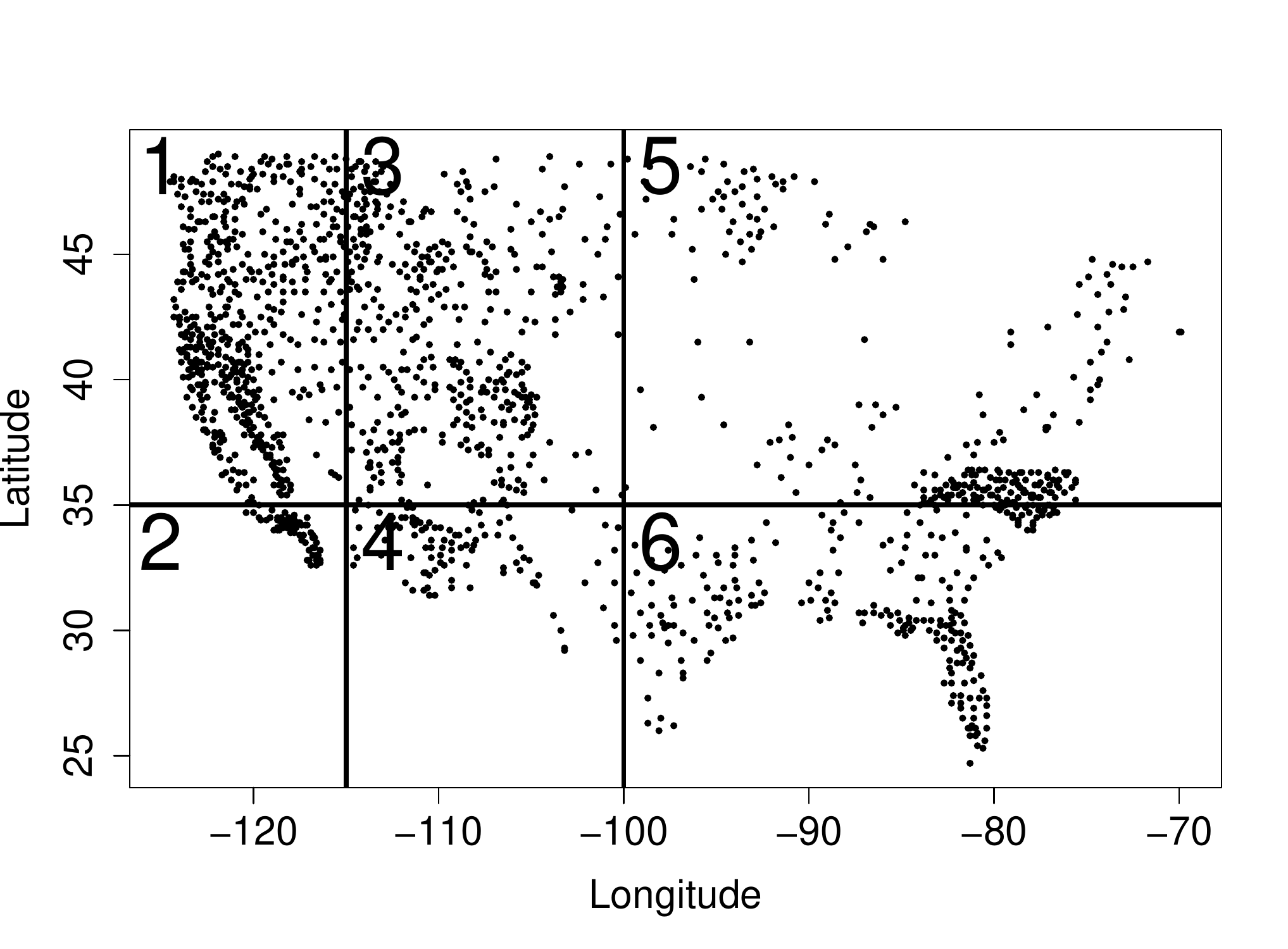}
\caption{Locations of the weather stations of the WFAS data and the 6 regions used in the simulation study visualized on the map.}
\label{regions}
\end{figure}

In this section we study if the proposed estimators of the tSAR model behave in a reasonable way and how they compare to the estimators of the already existing SAR model. 
\newline
\newline
We simulate from a tSAR model in the following way. 
\begin{enumerate}
\item (number of locations $n$) We specify the number of locations $n$ as 250 or 1500. 
\item (proximity matrix $W$) We use the first $n$ longitude/latitude values of the WFAS data set introduced in Section \ref{datadesc} to determine locations and corresponding neighborhoods. We set the proximity matrix $W$ equal to a nearest neighbors matrix with $k=30$ neighbors. 
\item (covariates $\boldsymbol x_1, \ldots, \boldsymbol x_7$)  We obtain the covariates $\boldsymbol x_1, \ldots, \boldsymbol x_7$ by sampling $n$ times independently from the following distributions:
\begin{equation}
\begin{split}
\boldsymbol x_1, \ldots, \boldsymbol x_5: & \text{ standard normal}\\ 
\boldsymbol x_6 \hspace*{0.5cm}:& \text{ bernoulli with }p=0.3\\
\boldsymbol x_7 \hspace*{0.5cm}:& \text{ bernoulli with }p=0.7
\end{split}
\end{equation}
\item (degrees of freedom $\nu$) We specify the degrees of freedom $\nu$ as 4 or 20. 
\item (simulation of \(\boldsymbol\epsilon\))  To account for a varying variance, we define 6 regions (see Figure \ref{regions}) with corresponding \(s_1 = 4, s_2 = 0.6, s_3 = 5, s_4 = 0.3, s_5 = 4, s_6 = 6\) and simulate independently for \(i = 1, \ldots, n \): If location \(i\) belongs to region \(j\) simulate \(\epsilon_{i}\) from \(t(0,s_j^2, \nu)\).

\item (coefficients $\boldsymbol\beta$) We set 
\begin{equation*}
\beta_0=3,~ \beta_1=10,~ \beta_2=4,~ \beta_3=5,~ \beta_4=2,~ \beta_5=8,~ \beta_6=1,~ \beta_7=3 .
\end{equation*}
\item (spatial parameter $\lambda$) We specify $\lambda$ as 0.4 or 0.8.
\item (response $\boldsymbol y$) According to the assumptions of the tSAR model we set 
  \begin{equation*}
\boldsymbol y = (\Id_n - \lambda W)^{-1} \boldsymbol\epsilon + X\boldsymbol\beta  ,
\end{equation*}
where $\boldsymbol\beta=(\beta_0,\beta_1,\ldots,\beta_7)^T$ and $X=(\boldsymbol 1^T, \boldsymbol x_1^T, \ldots, \boldsymbol x_7^T)$.
\end{enumerate}

Choosing between SAR and tSAR and the different choices for $\Sigma_{\epsilon}$ leads to 6 different models (see Table \ref{mod}) that are estimated from the simulated data.

\begin{table}[H]
\centering
\begin{tabular}{ccl}
model &  & $\Sigma_{\epsilon}$  \\
1&SAR&$\Id_n$\\
2&tSAR&$\Id_n$\\
3&SAR& local regression variance matrix\\
4&tSAR&local regression variance matrix\\
5&SAR&true\\
6&tSAR&true\\
\end{tabular}
\caption{Different models estimated in the simulation study.}
\label{mod}
\end{table} 

In the tSAR model we have one additional parameter $\nu$, the degrees of freedom, which was assumed to be known in Section \ref{tsar}. Instead of specifying this parameter we use numerical optimization to obtain an estimate for it. We use the \(\texttt{R}\) function \(\texttt{optimize}\) with high tolerance (tolerance = 1) to speed up computation. Here we allow $\nu$ to be a real parameter between 3 and 20. 

Note that in the SAR model \(\sigma\) is the standard deviation of \(\epsilon_i/\sqrt{(\Sigma_{\epsilon})_{ii}} \), whereas in the tSAR model \(\sigma\) is the square root of the scale parameter of \(\epsilon_i/\sqrt{(\Sigma_{\epsilon})_{ii}} \) and the standard deviation is given by \(\sqrt{\frac{\nu}{\nu - 2}} \sigma\). For easier comparison we introduce \emph{\(s\), the standard deviation of \(\epsilon_i/\sqrt{(\Sigma_{\epsilon})_{ii}} \)}, and define its estimate \(\hat s\), depending on the model, by
\begin{equation}
\begin{split}
\hat s & \colonequals \hat \sigma ~~~~~~~~~~~~~~~\text{ if the SAR model is used}, \\
\hat s & \colonequals \sqrt{\frac{\nu}{\nu - 2}} ~~\hat \sigma ~~\text{ if the tSAR model is used} .
\end{split}
\end{equation}

The results of the simulation study are shown in Table \ref{sumsim}. To evaluate the estimates we use the root mean squared error which is given by 
\begin{equation}
\text{RMSE}(\hat\theta) = \sqrt{\frac{1}{p}\sum_{j=1}^p\frac{1}{r}\sum_{i=1}^r (\theta_{j} - \hat{\theta}_{ji})^2} ,
\end{equation}
where $r$ is the number of replications (in our case $r=500$), $\theta_j$ is the $j$-th component of the $p$-dimensional vector $\boldsymbol \theta$ and $\hat{\theta}_{ji}$ its  estimate in the $i$-th replication.

First we analyze the results with respect to the number of locations $n$. We compare models that only differ in the choice of this parameter. One usually expects that the root mean squared error decreases as the number of stations increases. We observe this behavior for all parameters in cases where $\Sigma_{\epsilon}$ is the true value or the local regression variance matrix. If $\Sigma_{\epsilon}$ is the identity matrix this does not hold for the parameter $s$. The parameter $s$ scales $\Sigma_{\epsilon}$ and if $\Sigma_{\epsilon}$ is specified incorrectly we cannot expect a reasonable estimate for $s$. Furthermore, the results show that the choice of $\Sigma_{\epsilon}$ has an influence on the  estimates for $\boldsymbol \beta$. Comparing models that only differ in the choice of $\Sigma_{\epsilon}$, the best estimates are obtained when $\Sigma_{\epsilon}$ is the true value, the second best when $\Sigma_{\epsilon}$ is the local regression variance matrix and the worst when $\Sigma_{\epsilon} = \Id_n$. There is a notable difference between $\RMSE(\hat{\boldsymbol\beta})$ in cases where $\Sigma_{\epsilon} = \Id_n$ compared to cases where $\Sigma_{\epsilon}$ is equal to the local regression variance matrix. This shows that introducing the local regression variance matrix brings a notable improvement for estimating \(\boldsymbol\beta\) compared to the trivial choice $\Sigma_{\epsilon} = \Id_n$. For $\lambda$, reasonable estimates are provided in all cases whereas the best estimates are usually obtained when $\Sigma_{\epsilon}$ is the true value. 
We see that the choice of $\Sigma_{\epsilon}$ also has influence on the estimates for $s$, where the influence is similar as for $\boldsymbol\beta$, i.e., the best estimates are obtained when $\Sigma_{\epsilon}$ is the true value, the second best when $\Sigma_{\epsilon}$ is equal to the local regression variance matrix and the worst when $\Sigma_{\epsilon} = \Id_n$. The differences in $\RMSE(\hat s)$ are rather big since it is difficult to estimate $s$, which scales $\Sigma_{\epsilon}$, if  $\Sigma_{\epsilon}$ is not specified correctly. 
Analysing the estimation of $\nu$ we observe that big values of $\RMSE(\hat\nu)$ are obtained in cases where $\nu = 20$ and $\Sigma_{\epsilon}$ is not equal to the true value. In these cases $\nu$ was estimated too low. Specifying $\Sigma_{\epsilon}$ incorrectly causes that the variance of the residuals is estimated too low or too high for some of them which then causes that a $t$-distribution with lower degrees of freedom provides a better fit. 

\begin{table}[H]
\centering
\resizebox{1.1\columnwidth}{!}{%
\begin{tabular}{rrrrrrrrrrrrrrr}
\cline{3-6} 
 &  & \multicolumn{4}{|c|}{RMSE} &  &  &  &  &  &  &  &  \\ 
  \hline
  $n$ & model & $\hat{\boldsymbol\beta}$ & $\hat\lambda$ & $\hat s$ & $\hat\nu$ & $ll$ & $\hat{\boldsymbol\beta}$ & $\boldsymbol\beta$ & $\hat\lambda$ & $\lambda$ & $\hat s$ & $s$ & $\hat\nu$ & $\nu$ \\ 
250 & 1 & 0.428 & 0.086 & 2.961 &  & -722 & 4.49 & 4.5 & 0.31 & 0.4 & 4.38 & 1.41 &  &  \\ 
250 & 2 & 0.425 & 0.040 & 2.965 & 0.41 & -660 & 4.49 & 4.5 & 0.36 & 0.4 & 4.38 & 1.41 & 3.59 & 4 \\ 
250 & 3 & 0.106 & 0.062 & 0.390 &  & -567 & 4.50 & 4.5 & 0.34 & 0.4 & 1.02 & 1.41 &  &  \\ 
250 & 4 & 0.107 & 0.047 & 0.389 & 0.41 & -521 & 4.50 & 4.5 & 0.35 & 0.4 & 1.03 & 1.41 & 3.59 & 4 \\ 
250 & 5 & 0.069 & 0.032 & 0.032 &  & -474 & 4.50 & 4.5 & 0.37 & 0.4 & 1.38 & 1.41 &  &  \\ 
250 & 6 & 0.069 & 0.035 & 0.029 & 2.17 & -457 & 4.50 & 4.5 & 0.36 & 0.4 & 1.39 & 1.41 & 4.98 & 4 \\ 
250 & 1 & 0.651 & 0.054 & 2.972 &  & -728 & 4.51 & 4.5 & 0.75 & 0.8 & 4.39 & 1.41 &  &  \\ 
250 & 2 & 0.652 & 0.035 & 2.970 & 0.41 & -665 & 4.51 & 4.5 & 0.77 & 0.8 & 4.38 & 1.41 & 3.59 & 4 \\ 
250 & 3 & 0.165 & 0.049 & 0.474 &  & -576 & 4.50 & 4.5 & 0.75 & 0.8 & 0.94 & 1.41 &  &  \\ 
250 & 4 & 0.164 & 0.034 & 0.478 & 0.41 & -529 & 4.50 & 4.5 & 0.77 & 0.8 & 0.94 & 1.41 & 3.59 & 4 \\ 
250 & 5 & 0.113 & 0.022 & 0.035 &  & -480 & 4.50 & 4.5 & 0.78 & 0.8 & 1.38 & 1.41 &  &  \\ 
250 & 6 & 0.120 & 0.020 & 0.025 & 2.34 & -464 & 4.50 & 4.5 & 0.78 & 0.8 & 1.39 & 1.41 & 5.05 & 4 \\ 
250 & 1 & 0.323 & 0.067 & 2.220 &  & -652 & 4.49 & 4.5 & 0.33 & 0.4 & 3.27 & 1.05 &  &  \\ 
250 & 2 & 0.321 & 0.037 & 2.223 & 16.41 & -607 & 4.49 & 4.5 & 0.36 & 0.4 & 3.28 & 1.05 & 3.59 & 20 \\ 
250 & 3 & 0.073 & 0.048 & 0.101 &  & -485 & 4.50 & 4.5 & 0.35 & 0.4 & 0.95 & 1.05 &  &  \\ 
250 & 4 & 0.074 & 0.043 & 0.098 & 16.14 & -465 & 4.50 & 4.5 & 0.36 & 0.4 & 0.96 & 1.05 & 3.87 & 20 \\ 
250 & 5 & 0.052 & 0.032 & 0.022 &  & -402 & 4.50 & 4.5 & 0.37 & 0.4 & 1.03 & 1.05 &  &  \\ 
250 & 6 & 0.053 & 0.036 & 0.020 & 5.67 & -402 & 4.50 & 4.5 & 0.36 & 0.4 & 1.03 & 1.05 & 16.41 & 20 \\ 
250 & 1 & 0.462 & 0.047 & 2.219 &  & -657 & 4.50 & 4.5 & 0.75 & 0.8 & 3.27 & 1.05 &  &  \\ 
250 & 2 & 0.460 & 0.031 & 2.218 & 16.41 & -613 & 4.50 & 4.5 & 0.77 & 0.8 & 3.27 & 1.05 & 3.59 & 20 \\ 
250 & 3 & 0.111 & 0.045 & 0.170 &  & -499 & 4.50 & 4.5 & 0.76 & 0.8 & 0.88 & 1.05 &  &  \\ 
250 & 4 & 0.113 & 0.036 & 0.169 & 16.25 & -476 & 4.50 & 4.5 & 0.76 & 0.8 & 0.88 & 1.05 & 3.76 & 20 \\ 
250 & 5 & 0.080 & 0.021 & 0.022 &  & -408 & 4.50 & 4.5 & 0.78 & 0.8 & 1.03 & 1.05 &  &  \\ 
250 & 6 & 0.084 & 0.022 & 0.015 & 5.14 & -409 & 4.50 & 4.5 & 0.78 & 0.8 & 1.04 & 1.05 & 16.82 & 20 \\ 
1500 & 1 & 0.175 & 0.009 & 3.213 &  & -4430 & 4.50 & 4.5 & 0.39 & 0.4 & 4.63 & 1.41 &  &  \\ 
1500 & 2 & 0.175 & 0.005 & 3.214 & 0.41 & -4028 & 4.50 & 4.5 & 0.39 & 0.4 & 4.63 & 1.41 & 3.59 & 4 \\ 
1500 & 3 & 0.034 & 0.012 & 0.385 &  & -3162 & 4.50 & 4.5 & 0.39 & 0.4 & 1.03 & 1.41 &  &  \\ 
1500 & 4 & 0.034 & 0.009 & 0.386 & 0.39 & -2986 & 4.50 & 4.5 & 0.39 & 0.4 & 1.03 & 1.41 & 3.64 & 4 \\ 
1500 & 5 & 0.029 & 0.004 & 0.009 &  & -2981 & 4.50 & 4.5 & 0.40 & 0.4 & 1.40 & 1.41 &  &  \\ 
1500 & 6 & 0.029 & 0.008 & 0.008 & 0.51 & -2865 & 4.50 & 4.5 & 0.39 & 0.4 & 1.41 & 1.41 & 4.18 & 4 \\ 
1500 & 1 & 0.274 & 0.012 & 3.237 &  & -4464 & 4.50 & 4.5 & 0.79 & 0.8 & 4.65 & 1.41 &  &  \\ 
1500 & 2 & 0.274 & 0.009 & 3.239 & 0.41 & -4060 & 4.50 & 4.5 & 0.79 & 0.8 & 4.65 & 1.41 & 3.59 & 4 \\ 
1500 & 3 & 0.051 & 0.010 & 0.446 &  & -3205 & 4.50 & 4.5 & 0.79 & 0.8 & 0.97 & 1.41 &  &  \\ 
1500 & 4 & 0.052 & 0.008 & 0.445 & 0.40 & -3027 & 4.50 & 4.5 & 0.79 & 0.8 & 0.97 & 1.41 & 3.63 & 4 \\ 
1500 & 5 & 0.043 & 0.005 & 0.001 &  & -3021 & 4.50 & 4.5 & 0.80 & 0.8 & 1.42 & 1.41 &  &  \\ 
1500 & 6 & 0.044 & 0.006 & 0.006 & 0.50 & -2902 & 4.50 & 4.5 & 0.79 & 0.8 & 1.42 & 1.41 & 4.15 & 4 \\ 
1500 & 1 & 0.134 & 0.012 & 2.407 &  & -3997 & 4.50 & 4.5 & 0.39 & 0.4 & 3.46 & 1.05 &  &  \\ 
1500 & 2 & 0.134 & 0.005 & 2.407 & 16.41 & -3722 & 4.50 & 4.5 & 0.39 & 0.4 & 3.46 & 1.05 & 3.59 & 20 \\ 
1500 & 3 & 0.024 & 0.015 & 0.081 &  & -2694 & 4.50 & 4.5 & 0.38 & 0.4 & 0.97 & 1.05 &  &  \\ 
1500 & 4 & 0.024 & 0.014 & 0.082 & 11.97 & -2669 & 4.50 & 4.5 & 0.39 & 0.4 & 0.97 & 1.05 & 8.18 & 20 \\ 
1500 & 5 & 0.022 & 0.005 & 0.003 &  & -2547 & 4.50 & 4.5 & 0.40 & 0.4 & 1.05 & 1.05 &  &  \\ 
1500& 6 & 0.022 & 0.009 & 0.002 & 3.67 & -2545 & 4.50 & 4.5 & 0.39 & 0.4 & 1.05 & 1.05 & 17.56 & 20 \\ 
1500 & 1 & 0.205 & 0.005 & 2.406 &  & -4028 & 4.51 & 4.5 & 0.79 & 0.8 & 3.46 & 1.05 &  &  \\ 
1500 & 2 & 0.204 & 0.003 & 2.408 & 16.41 & -3753 & 4.51 & 4.5 & 0.80 & 0.8 & 3.46 & 1.05 & 3.59 & 20 \\ 
1500 & 3 & 0.036 & 0.009 & 0.136 &  & -2737 & 4.50 & 4.5 & 0.79 & 0.8 & 0.92 & 1.05 &  &  \\ 
1500 & 4 & 0.036 & 0.008 & 0.135 & 12.42 & -2713 & 4.50 & 4.5 & 0.79 & 0.8 & 0.92 & 1.05 & 7.69 & 20 \\ 
1500& 5 & 0.032 & 0.003 & 0.003 &  & -2578 & 4.50 & 4.5 & 0.80 & 0.8 & 1.05 & 1.05 &  &  \\ 
1500 & 6 & 0.032 & 0.004 & 0.001 & 3.70 & -2580 & 4.50 & 4.5 & 0.80 & 0.8 & 1.05 & 1.05 & 17.58 & 20 \\ 
   \hline
\end{tabular}%
}
\caption{Results of the simulation study. The first two columns specify the number of locations and the model. Other columns show the root mean squared error, the average log-likelihood ($ll$), the true parameter and the average of estimated parameters. For $\boldsymbol \beta$ we average over its components.}
\label{sumsim}
\end{table}

Evaluating the overall fit with the log-likelihood and comparing models that only differ in the choice of one parameter we see that the choice between SAR and tSAR and the choice of $\Sigma_{\epsilon}$ has influence. The tSAR model leads to higher likelihood values when $\nu=4$ or mostly similar values when $\nu =20$. For $\Sigma_{\epsilon}$ the highest likelihood values are obtained when $\Sigma_{\epsilon}$ is the true value, the second highest when $\Sigma_{\epsilon}$ is equal to the local regression variance matrix and the lowest when $\Sigma_{\epsilon} = \Id_n$.

\section{Application}
\label{application}

We use the two models, SAR and tSAR, to fit data to assess the risk of fire danger in the US. 

\subsection{Data description}
\label{datadesc}
The data is obtained from the Wildland Fire Assessment System (WFAS) and contains the following variables observed at 1542 stations on the 23rd of June 2015.
\vspace*{0.4cm}
\begin{itemize}
\item \(Elev\) = Elevation in feet divided by 100
\item \(Lat\) = Latitude
\item \(Long\) = Longitude
\item \(Tmp\) = Temperature in Fahrenheit
\item \(RH\) = Relative humidity in percent
\item \(Wind\) = Wind speed (10 min avg wind) in mi/h
\item \(PPT\) = 24h precipitation in inches
\item \(BI\) = Burning Index calculated according to the National Fire Danger Rating System (cf., National Wildfire Coordinating Group\cite{nfdrs2002selfstudy}) (number related to the contribution of fire behavior to the effort of containing a fire. It is expressed as a numeric value closely related to the flame length in feet multiplied by 10.)
\end{itemize}

\subsection{Model fitting}
We consider the Burning Index \(BI\) as response variable and the other variables as covariates. These covariates can be measured using simple weather station technology. For our approach, there is no expert knowledge required compared to the calculation of the Burning Index according to the National Fire Danger Rating System.

\begin{figure}[H]
\centerline{%
\includegraphics[width=0.5\textwidth]{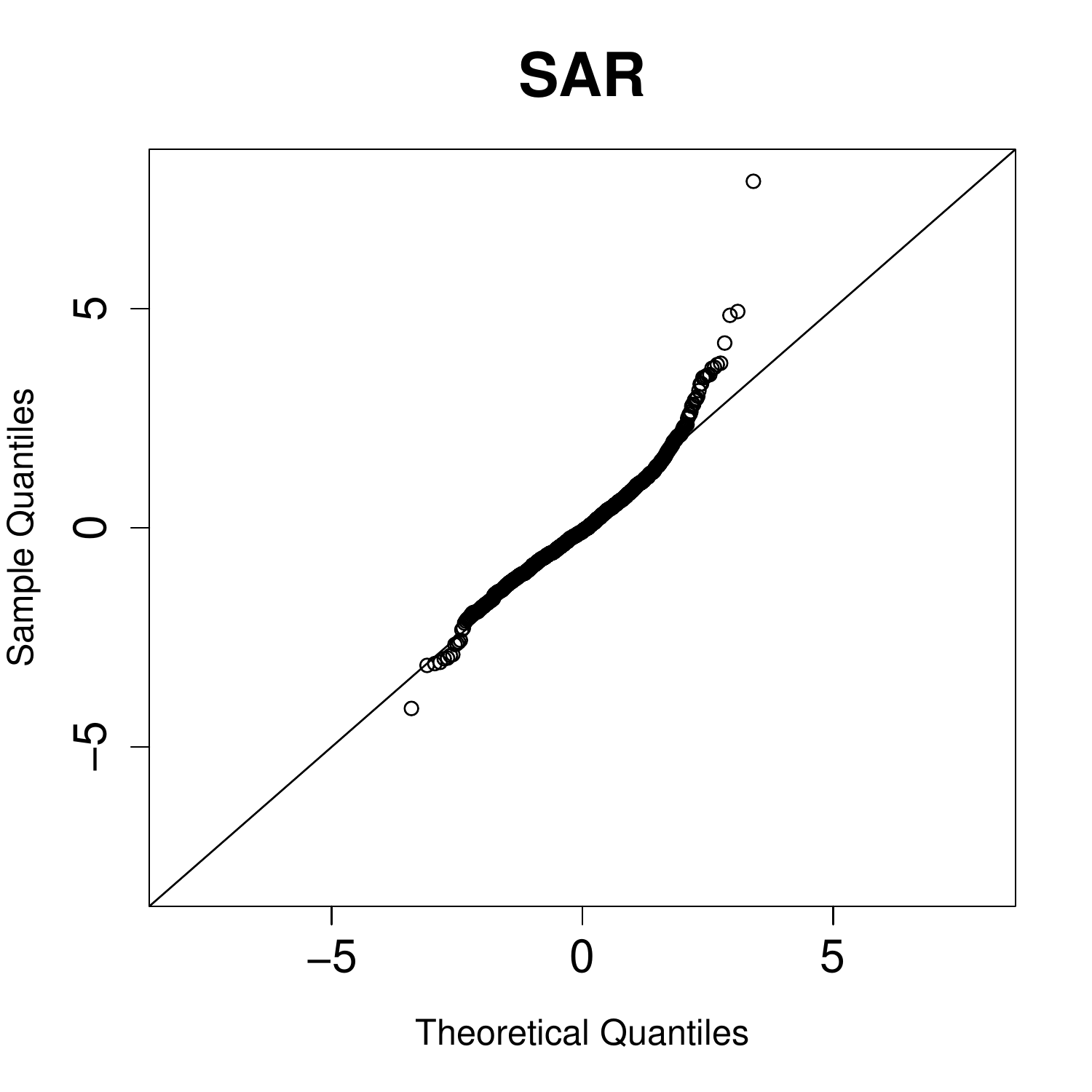}%
\includegraphics[width=0.5\textwidth]{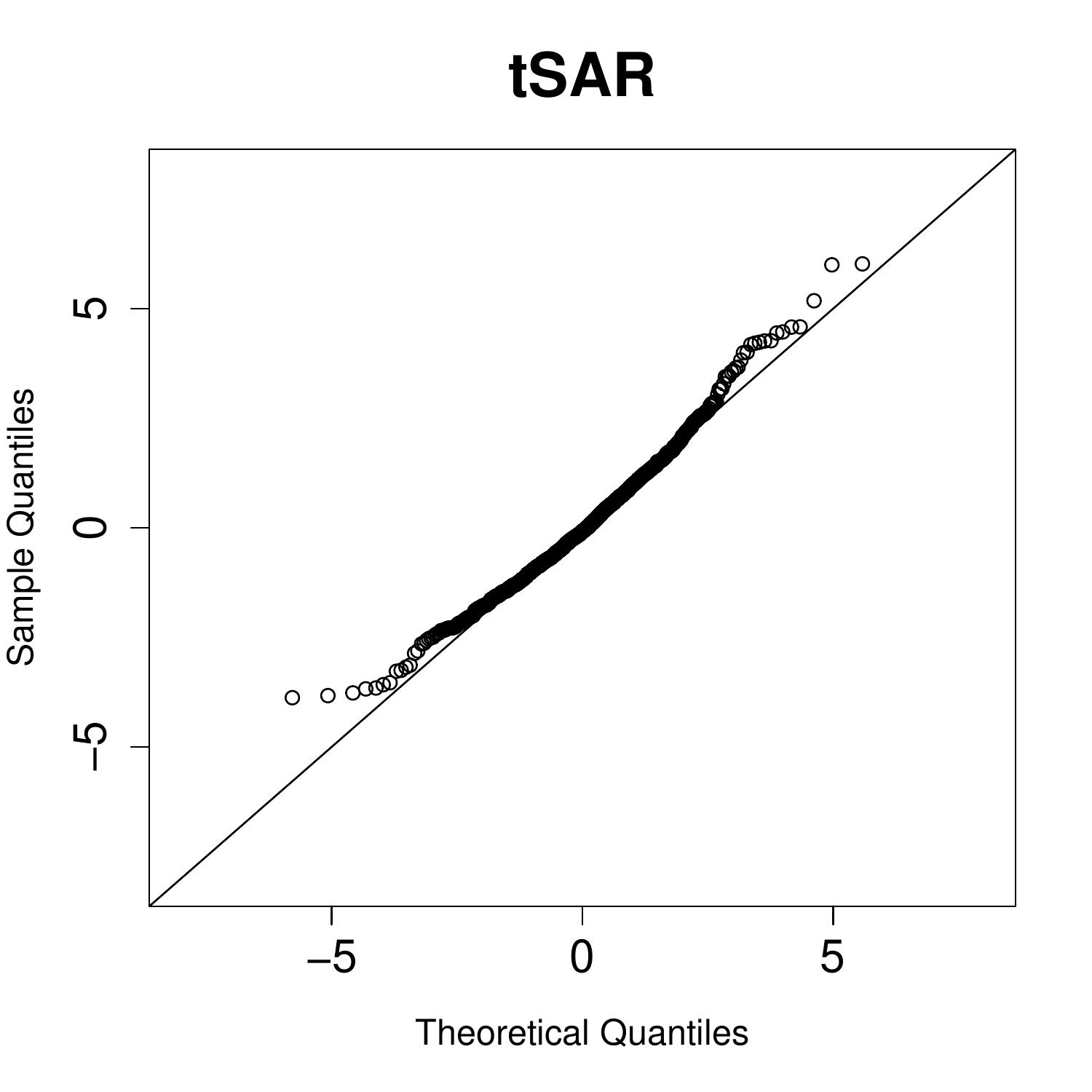}%
}%
\caption{qq-plots for a SAR and a tSAR model. We plot the quantiles of the standard normal distribution against the quantiles of the standardized residuals of the SAR model and the quantiles of the $t$-distribution with mean zero, scale parameter 1 and $6$ degrees of freedom against the quantiles of the standardized residuals of the tSAR model with $\nu = 6$.}
\label{bres}
\end{figure}

Fitting several SAR and tSAR models, we observed  misbehavior in the residuals. The residuals did not follow the desired normal or $t$-distribution. Figure \ref{bres} illustrates this problem for one case where we fit one SAR and one tSAR model with $\nu = 6$ degrees of freedom. We use $BI$ as response and all other variables as covariates. As proximity matrix $W$ we choose a nearest neighbors matrix with $k=30$ neighbors and for $\Sigma_{\epsilon}$ we use the local regression variance matrix of $Y$ with respect to $W$. 

To deal with this problem and to further improve our fit, we now consider Box-Cox transformations of the response variable (cf., Box and Cox\cite{box1964analysis}) for SAR models.  
We show how Box-Cox transformations that were developed for linear regression models can be used for  SAR and tSAR models. We are given \(\boldsymbol y =(y_1, \ldots, y_n)^T\), an observation of the random vector \(\boldsymbol Y = (Y_1, \ldots, Y_n)^T\).
For \(l \in \mathbb{R}\) and \(m \in \mathbb{R}\) such that \(Y_i > -m\) for all \(i = 1, \ldots,n\), the Box-Cox transformed variable \(Y_i^{m,l}\) is given by
\begin{equation}
Y_i^{m,l} = \left\{\begin{array}{cl} \frac{(Y_i + m)^{l} - 1}{l}, & \mbox{if } l \neq 0     \\ \log(Y_i + m), & \mbox{else} \end{array}\right.  .
\label{transfo}
\end{equation}
We consider \(l\) and \(m\) fixed and assume that \(\boldsymbol Y^{m,l}\) is distributed according to a SAR or tSAR model with parameters \(\boldsymbol \theta^{m,l} = (\boldsymbol\beta^{m,l},\sigma^{m,l},\lambda^{m,l})\). We denote its log-likelihood by \(\ell_{S}(\boldsymbol y^{m,l}|\boldsymbol \theta^{m,l})\) where the observation $ y_i^{m,l}$ of $ Y_i^{m,l}$ is obtained by applying the same transformation \ref{transfo} on the observation $y_i$. The density of $\boldsymbol Y$ can be obtained using the density transformation rule. The log-likelihood of \(\boldsymbol \theta^{m,l}\) with respect to the observations \(y_1, \dots, y_n\) is then given by
\begin{equation*}
\ell(\boldsymbol y|\boldsymbol \theta^{m,l}) =\sum_{i=1}^n (l - 1) \log(y_i + m) + \ell_{S}(\boldsymbol y^{m,l}|\boldsymbol \theta^{m,l}).
\end{equation*}
The log-likelihood is a sum of two components where the first component is independent of \(\boldsymbol \theta^{m,l}\), and therefore not needed for the maximization with regard to \(\boldsymbol \theta^{m,l}\). So we need to maximize the second component which we know how to do since it is the log-likelihood of a SAR or tSAR model. Knowing the log-likelihood, the corresponding BIC is 
\begin{equation*}
\BIC(\boldsymbol y, \boldsymbol \theta^{m,l}) = -2 \ell(\boldsymbol y|\boldsymbol \theta^{m,l}) + \dim(\boldsymbol \theta^{m,l}) \log(n),
\end{equation*}
which can be used for selection among different models corresponding to different \(m\) and \(l\) values.

For fitting SAR models we use a step wise procedure where we adjust the Box-Cox transformation parameter and eliminate a non-significant covariate in each step. The procedure (Algorithm \ref{algo}) for a given variable \(R\), parameter $m$ and proximity matrix \(W\) is shown in the following. The available covariates are denoted by $x_1, \ldots, x_p$.

\begin{algorithm}[H]
\caption{Step wise procedure for SAR models}
\label{algo}
\begin{algorithmic}[1]
\STATE \(\mathcal{X} \gets \{x_1, \ldots, x_p\}\)
\STATE $ maxp\gets 1 $ 
\STATE \(c \gets \{\} \)
\WHILE{\(maxp > 0.05\)}
\STATE \(\mathcal{X} = \mathcal{X} \setminus \{c\}\)
\FOR{\(l = -2,-1,-1/2,-1/3,0,1/3,1/2,1,2\)}
\STATE \(Y \gets \left\{\begin{array}{cl} \frac{(R + m)^{l} - 1}{l}, & \mbox{if } l \neq 0     \\ \log(R + m), & \mbox{else} \end{array}\right.   \)
\STATE \(mod_{l} \gets\) fitted SAR model with response variable \(Y\), covariates \(\mathcal{X}\), proximity matrix \(W\) and \(\Sigma_{\epsilon}\) is the local regression variance matrix of \(Y\) with respect to \(W\).
\ENDFOR
\STATE \(mod \gets\)  model with lowest BIC among \(\{ mod_{l}|l = -2,-1,-1/2,-1/3,0,1/3,1/2,1,2 \}\) 
\STATE \(maxp \gets\)  maximum of the p-values of the tests for significance of the coefficients in model \(mod\) 
\STATE \(c \gets\)  covariate corresponding to \(maxp\)
\ENDWHILE
\end{algorithmic}
\end{algorithm}

 Algorithm \ref{algo} is applied to the response variable \(BI\) with $m=10$ and different choices of the proximity matrix \(W\). Instead of iterating over different values for $m$ we choose one value, 10, to reduce computational time. For the proximity matrix we use nearest neighbors matrices with \(k=10, 20, 30, 40, 50\) neighbors and radius matrices with radius \(r=350,500\). So we obtain 7 different models corresponding to different proximity matrices. 

After fitting SAR models using the procedure just described, we fit tSAR models. We proceed in the following way. For a certain proximity matrix \(W\) we take the same covariates and transformation as in the corresponding just fitted SAR model and fit a tSAR model where we optimize the degrees of freedom parameter $\nu$ numerically. For the matrix \(\Sigma_{\epsilon}\) we use as before the local regression variance matrix of the transformed response variable with respect to \(W\). Table \ref{bibic} shows the BIC values of the models. If we consider only nearest neighbors matrices, we see that the BIC of the worst tSAR model is still lower than the BIC of the best SAR model. The best model is a tSAR model where the proximity matrix is a nearest neighbors  matrix with $k = 20$ neighbors. Estimates for this model are given in Table \ref{esttsar}.

\begin{table}[H]
\centering
\resizebox{1\columnwidth}{!}{%

\begin{tabular}{l|rrrrr|rr}
  \hline
model type & nn10 & nn20 & nn30 & nn40 & nn50 & r350 & r500 \\ 
  \hline
SAR & 12624.72 & 12488.34 & 12480.19 & 12499.37 & 12539.93 & 12617.03 & 12737.47 \\
tSAR &  12424.78 & \textbf{12400.30} & 12418.51 & 12438.61 & 12470.69 & 12561.87 & 12684.05 \\
$l$ & 1/3 & 1/3 & 1/3 & 1/3 & 1/3 & 1/3 & 1/3 \\
\end{tabular}%
}
\caption{BIC for different models for transformed \(BI\) and the value of the transformation parameter $l$. ``nnx" means that a nearest neighbors matrix with x neighbors was used and ``rx" means that a radius matrix with radius x was used.}
\label{bibic}
\end{table}

\begin{table}[H]
\centering
\begin{tabular}{rrrr}
  \hline
 & estimate & $\hat{\se}$ & estimate$/\hat{\se}$\\ 
  \hline
Intercept & 7.72 & 1.14 & 6.76 \\ 
Elev & 0.01 & 0.00 & 2.85 \\ 
Lat & -0.07 & 0.03 & -2.44 \\ 
Long & -0.03 & 0.01 & -2.56 \\ 
RH & -0.04 & 0.00 & -12.39 \\ 
Wind & 0.16 & 0.01 & 20.97 \\ 
PPT & -0.80 & 0.13 & -6.34 \\
$\lambda$ & 0.85 &  &  \\ 
  $\sigma$ & 0.84 &  &  \\ 
  $\nu$ & 6.34 &  &  \\ 
   \hline

\end{tabular}
\caption{Parameter estimates, estimated standard errors and their quotient for the model with the best BIC of the Box-Cox transformed Burning Index $(m=10, l=\frac{1}{3})$. }
\label{esttsar}
\end{table}

\begin{figure}[H]
\centerline{%
\includegraphics[width=0.5\textwidth]{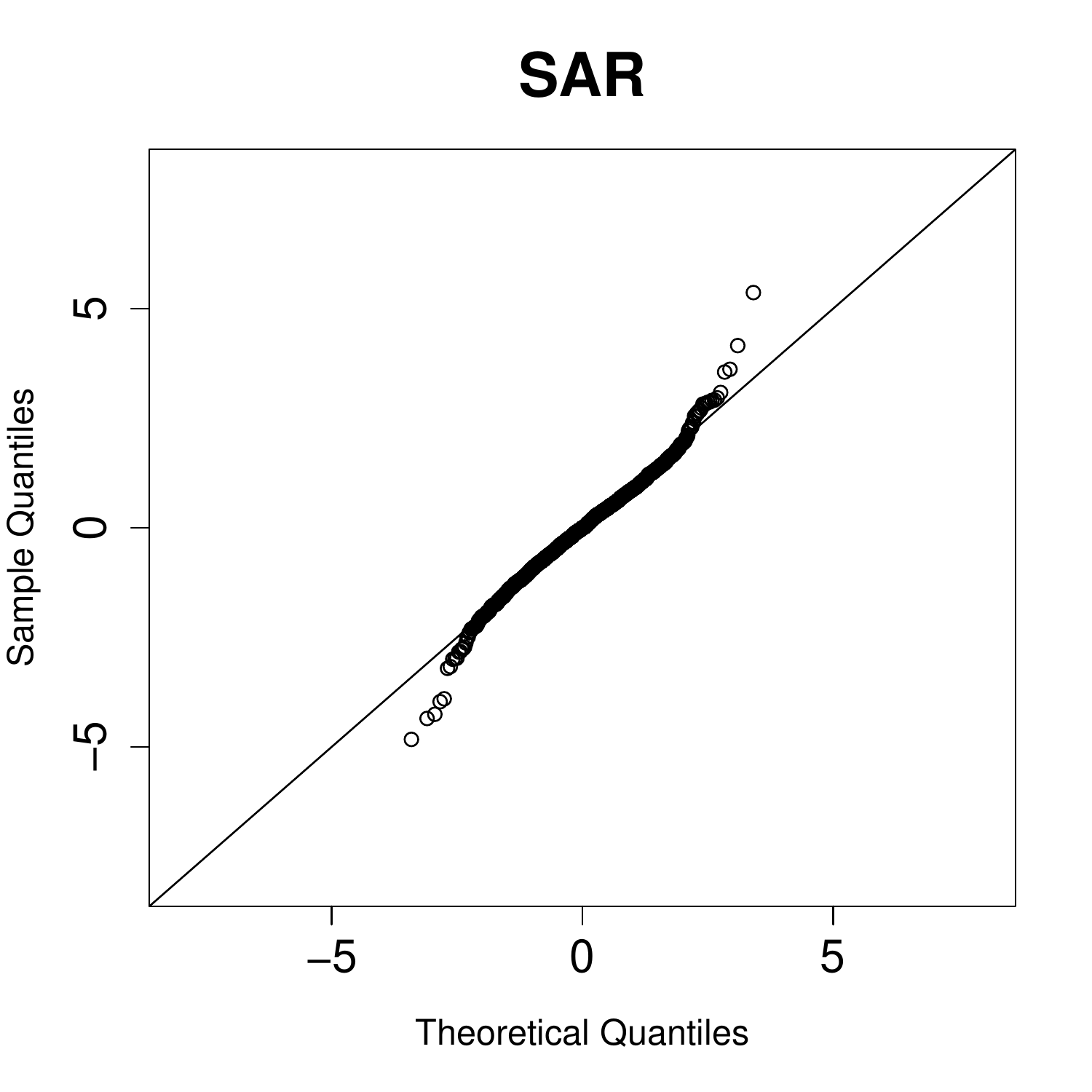}%
\includegraphics[width=0.5\textwidth]{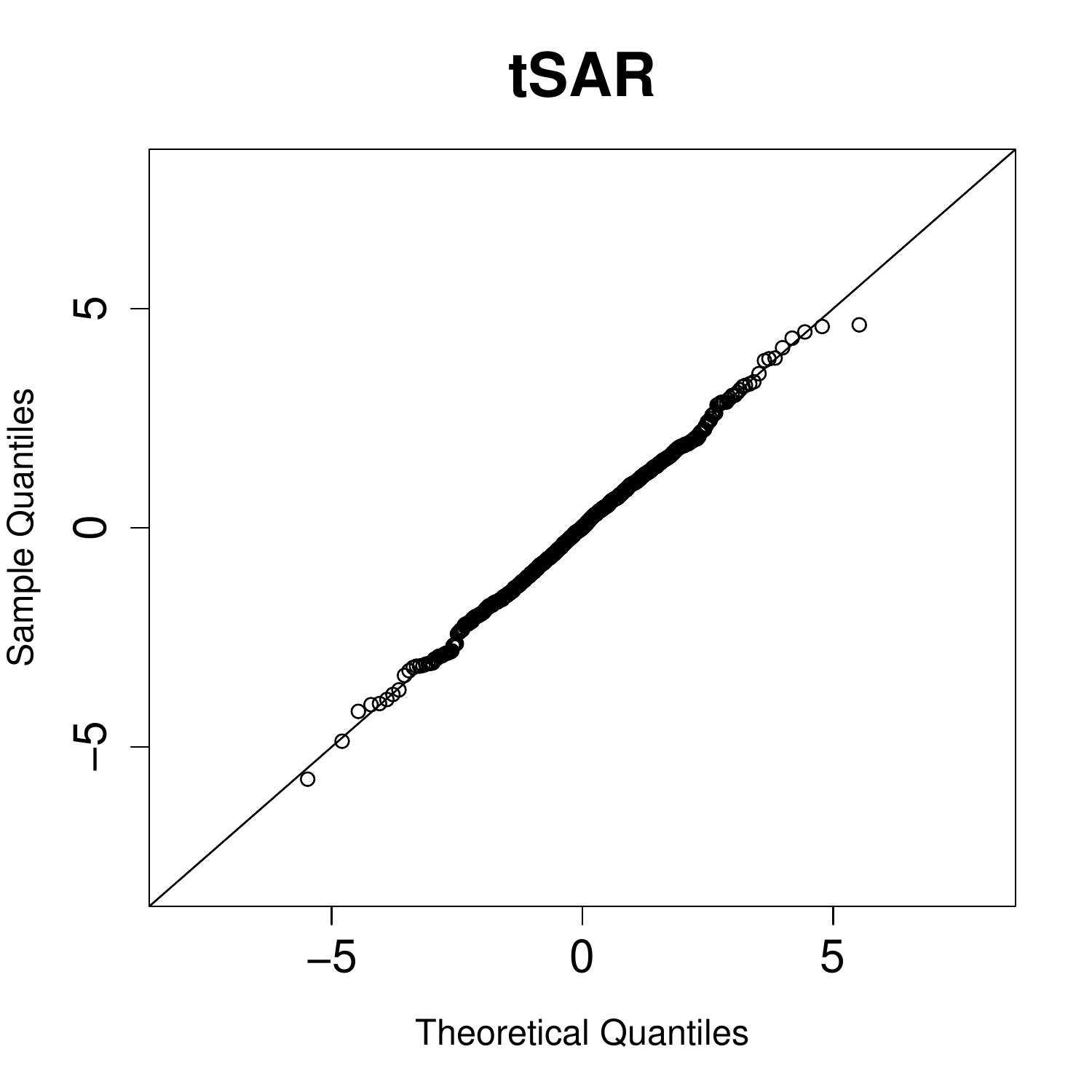}%
}%
\caption{qq-plots for the SAR and tSAR model with the best BIC value. }
\label{tsarresplot}
\end{figure}

In Figure \ref{tsarresplot} we check if the residuals of the best tSAR model have the distribution as expected. As the data points do not deviate far from the $x=y$ line, our fitted model seems to be appropriate. For comparison we also show this plot for the SAR model with the lowest BIC. We see that the tSAR model is not only preferred in terms of BIC.

\subsection{Out of sample prediction}
Now we perform out of sample predictions. This allows us to predict the Burning Index at locations where only the covariates are available. To do so, we need to relate a random variable at a location which was not part of the sample to $\boldsymbol Y$, the vector of random variables in the sample. For an out of sample random variable $Y_o$ at location $l_o$ we assume that
\begin{equation*}
Y_o = \boldsymbol \beta^T \boldsymbol x_o + \lambda \sum_{j \in N_o} w_{oj} (Y_j - \boldsymbol \beta^T \boldsymbol x_j ) + \epsilon_o ,
\end{equation*}
where $w_{oj}$ relates location $l_o$ to $l_j$ for $j=1...n$ such that $\sum_{j=1}^n w_{oj} =1$ to stay consistent with the row-standardized proximity matrix. We will choose $w_{oj}$ similar to how we chose the entries of the proximity matrix. If $W$ is a $k$ nearest neighbors matrix, $w_{oj}$ is the inverse distance between location $l_o$ and $l_j$ times a standardization constant, if location $l_j$ is among the $k$ nearest neighbors of $l_o$  and zero else. $N_o$ is the neighborhood of location $l_o$ defined as in Section \ref{sarmodel}.
For the error we assume \(\epsilon_o \sim N(0,\sigma^2 \Sigma_{o})\) in the case of a SAR model or \(\epsilon_o \sim t(0,\sigma^2 \Sigma_{o}, \nu)\) in the case of a tSAR model. Similar to the SAR and tSAR model, $\Sigma_{o}$ is assumed to be known. We specify $\Sigma_{o}$ similar to how we specified $\Sigma_{\epsilon}$. If $\Sigma_{\epsilon}$ is the local regression variance matrix of $\boldsymbol Y$, the diagonal entries of $\Sigma_{\epsilon}$  were calculated with linear regression residuals $r_1, \ldots r_n$. $\Sigma_{o}$ is then the empirical variance of $\{r_j|j \in N_o\}$.

With this assumption the expectation of $Y_o$ given $\boldsymbol Y$ is given by 
\begin{equation*}
\E(Y_o|\boldsymbol Y) = \E(Y_o|Y_j = y_j, j \in N_o) = \boldsymbol \beta^T \boldsymbol x_0 + \lambda \sum_{j \in N_o} w_{oj} (y_j - \boldsymbol \beta^T \boldsymbol x_j ),
\end{equation*}
where $\boldsymbol \beta$, $\sigma$ and $\nu$ are the parameters of the SAR or tSAR model for $\boldsymbol Y$.
So we define the local prediction of $Y_o$, where the neighbors' values are observed, by
\begin{equation*}
\hat y_{o|N_o} \colonequals \hat{\boldsymbol \beta}^T \boldsymbol x_o + \hat\lambda \sum_{j \in N_o} w_{oj} (y_j - \hat{\boldsymbol \beta}^T \boldsymbol x_j ) ,
\end{equation*}
where $\hat{\boldsymbol \beta}$ and $\hat\lambda$ are the estimates of the model for $\boldsymbol Y$. In addition to the prediction we provide confidence intervals. The $1 - \alpha$ confidence interval is given by
\begin{equation*}
\CI(1-\alpha) = 
\begin{cases}
\hat y_{o|N_o} \pm \Phi^{-1}(1-\frac{\alpha}{2},0,\hat\sigma^2 \Sigma_{o}) \text{ ~~for SAR} \\
\hat y_{o|N_o} \pm t^{-1}(1-\frac{\alpha}{2},0,\hat\sigma^2 \Sigma_{o},\nu) \text{ ~~for tSAR} \\
\end{cases},
\end{equation*}
where $\Phi^{-1}(1-\frac{\alpha}{2},0,\hat\sigma^2 \Sigma_{o})$ and $t^{-1}(1-\frac{\alpha}{2},0,\hat\sigma^2 \Sigma_{o},\nu)$ are the $1-\frac{\alpha}{2}$ quantiles of the $N(0,\hat\sigma^2 \Sigma_{o})$ and the $t(0,\hat\sigma^2 \Sigma_{o},\nu)$ distribution.

\begin{figure}[H]
\centerline{%
\includegraphics[width=0.5\textwidth]{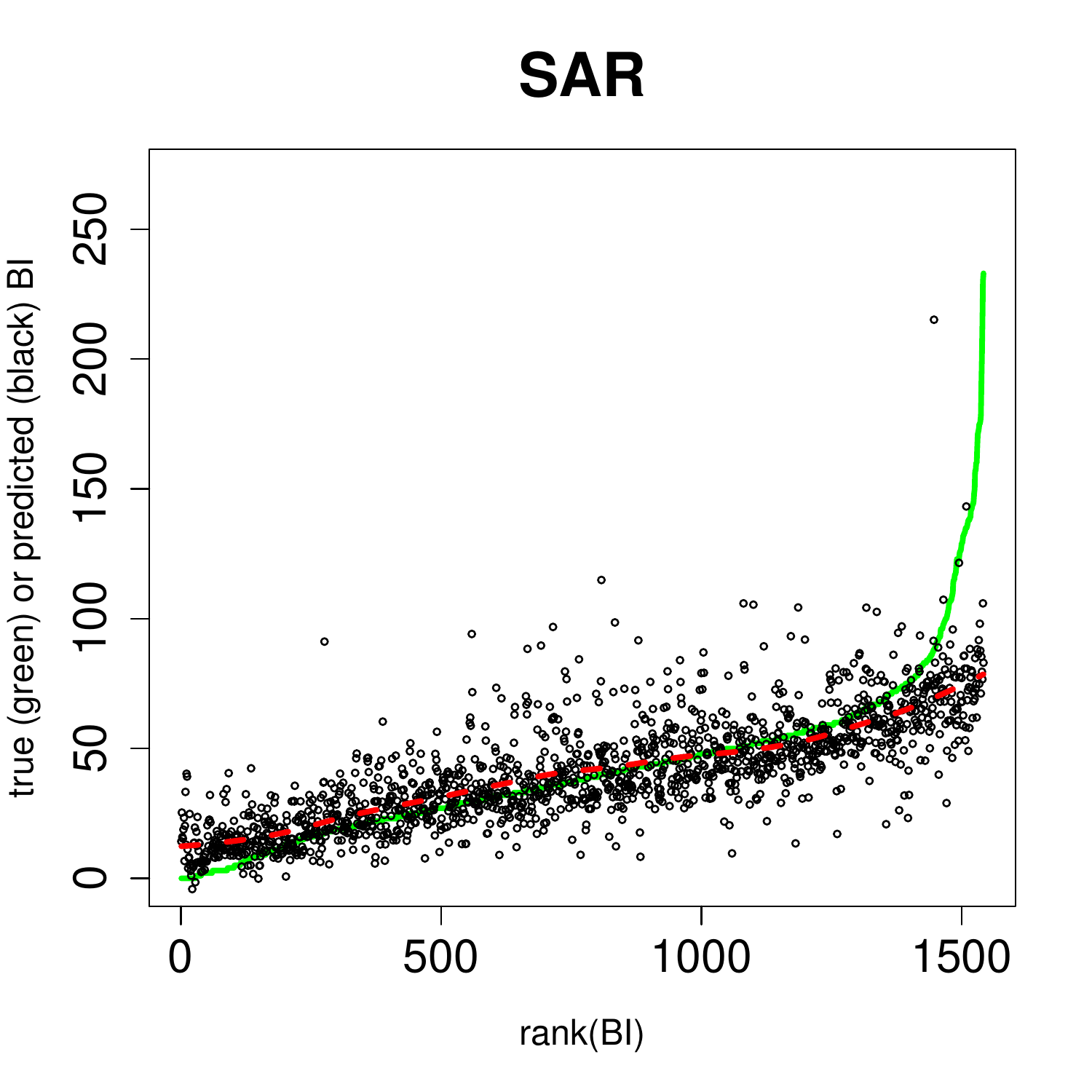}%
\includegraphics[width=0.5\textwidth]{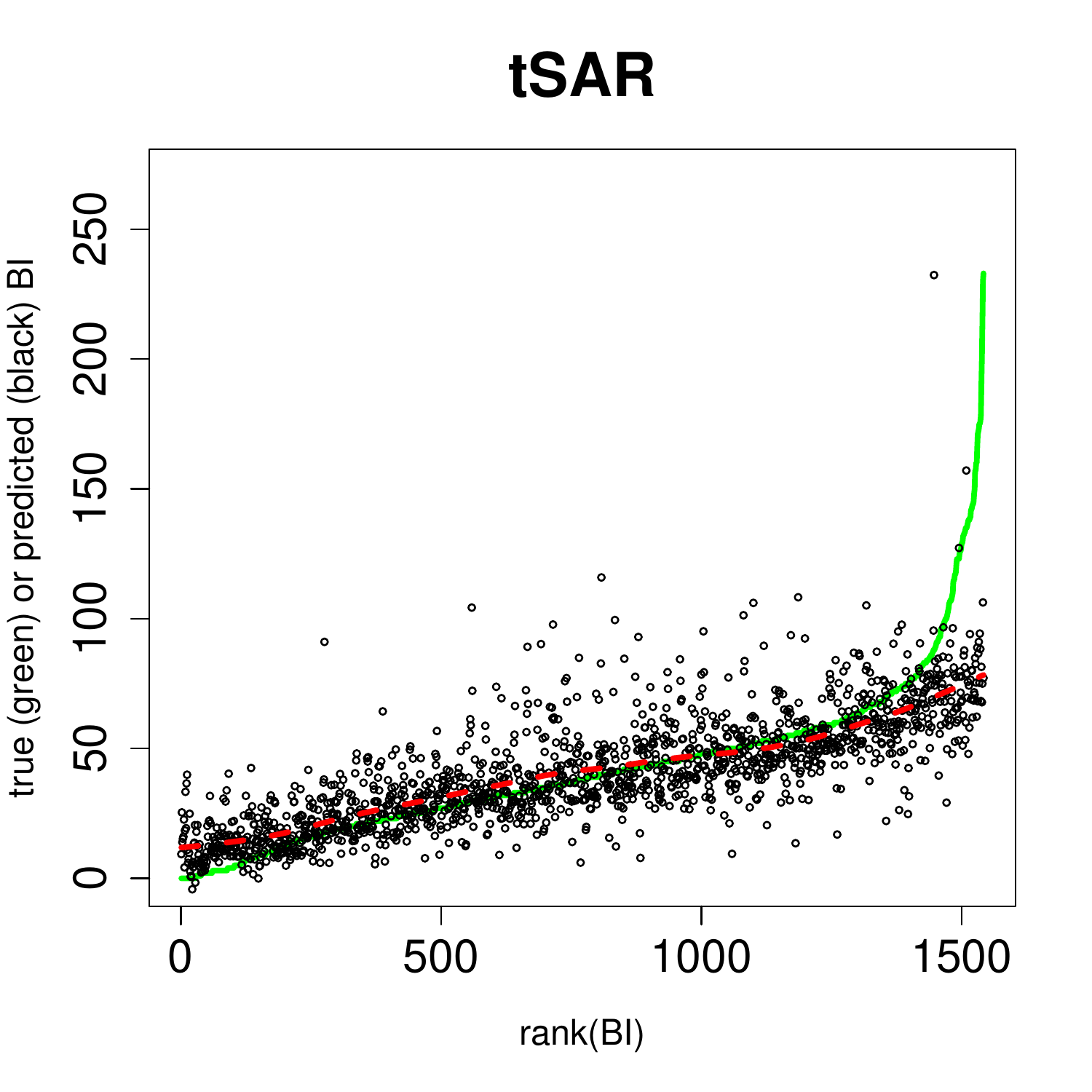}%
}%
\caption{True vs predicted Burning Index (BI). A smoothed curve for the predicted Burning Index was added in red. For better visualization the Burning Index was ordered.}
\label{tvsp}
\end{figure}

\begin{figure}[H]
\centerline{%
\includegraphics[width=0.5\textwidth]{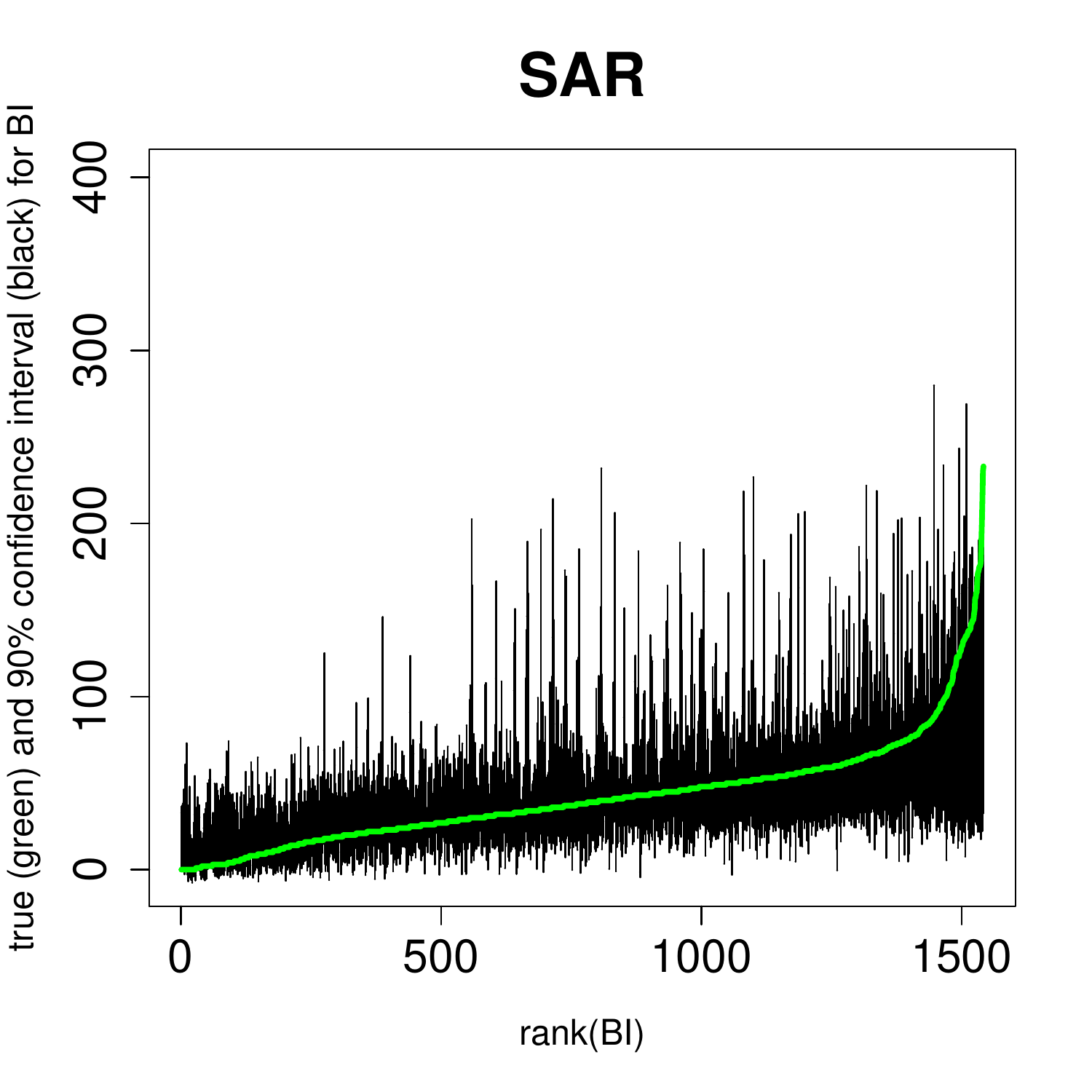}%
\includegraphics[width=0.5\textwidth]{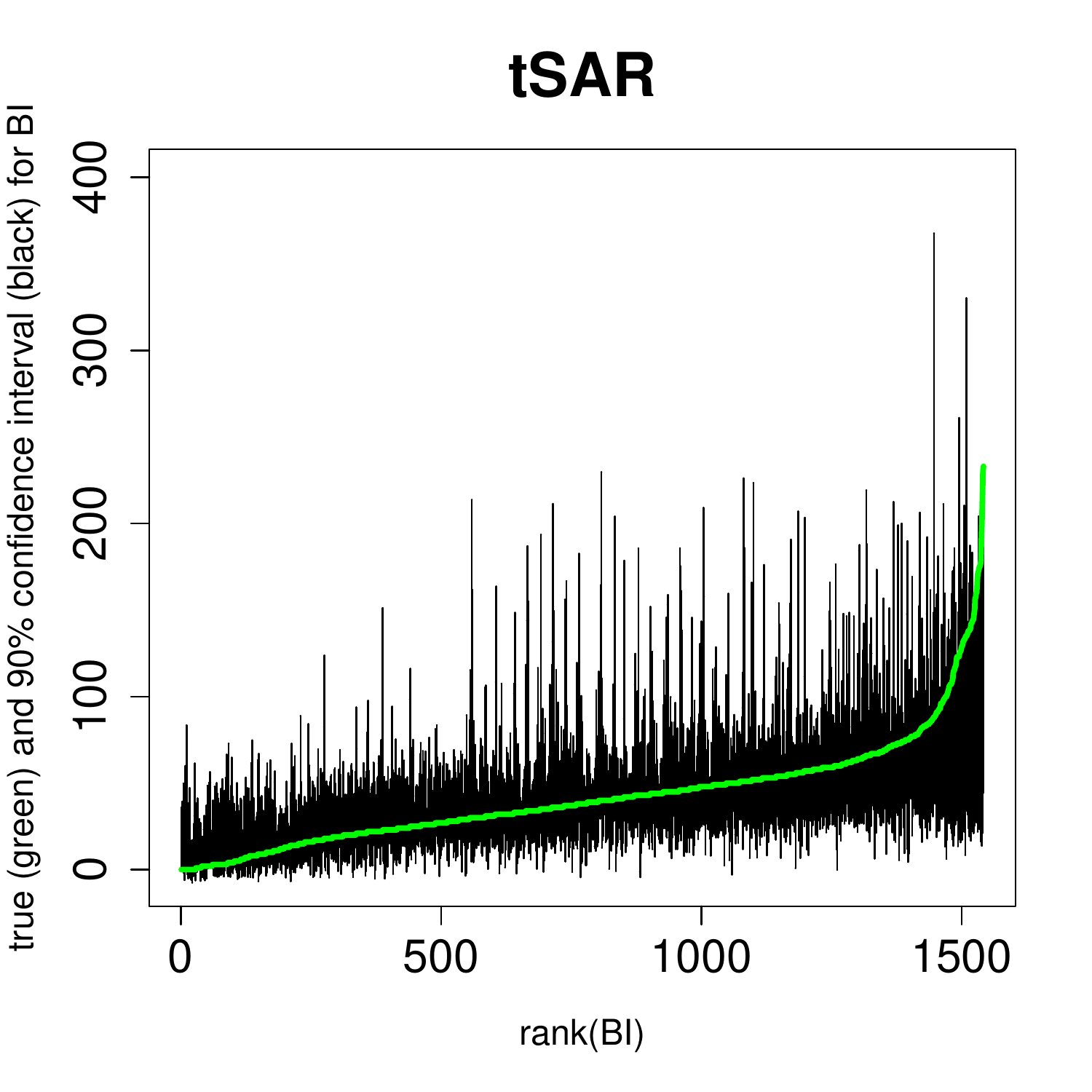}%
}%
\caption{Burning Index (BI) and its $90\%$ confidence intervals. For better visualization the Burning Index was ordered.}
\label{ci}
\end{figure}

To perform out of sample prediction, we divide our data set in 10 distinct batches. We use 9 batches for fitting the model and apply the same procedure as before. Our fitted model is the one with the lowest BIC. For the remaining batch data we perform out of sample prediction. Doing this 10 times gives us an out of sample prediction for every location. In every case the fitted model was a tSAR model. For comparison we also take the best SAR model for every case and perform out of sample prediction with this model. The predictions are shown in Figure \ref{tvsp} where we see that there is not a big difference between the SAR and the tSAR model. The prediction is influenced by the estimation of $\lambda$ and $\boldsymbol\beta$ where the SAR and the tSAR model provide similar estimates. The two models differ in the specification of the error distribution which influences confidence intervals. Figure \ref{ci} shows the confidence intervals and Table \ref{citable} the proportion of data points inside the corresponding confidence interval. We see that, in all three cases of confidence levels, this proportion is closer to the theoretical  confidence level for tSAR based confidence intervals.  To support this statement we conduct a likelihood ratio test (see Wilks \cite{wilks1938large}) for binomial data. We consider a theoretical confidence level of $1- \alpha$. Then we test the null hypothesis that the number of points lying outside the confidence interval is binomial distributed with success probability $\alpha$ against the alternative that it is binomial distributed with a success probability different than $\alpha$. The results of this test are shown in Table \ref{pval}. We see that higher $p$-values are obtained when the tSAR model is used. For the $99\%$ confidence interval the SAR model leads to a very small $p$-value and the null hypothesis is rejected at the $0.1 \%$ level. This can be explained by the fact that the normal distribution is not a good choice to model heavy tailed data.

\begin{table}[H]
\centering
\begin{tabular}{rrr}
  \hline
 & SAR & tSAR \\ 
  \hline
90$\%$ & 91.05$\%$ & 90.21$\%$\\ 
95$\%$ &  94.36$\%$ & 94.55$\%$ \\ 
99$\%$ & 97.93$\%$ & 98.96$\%$ \\  
   \hline
\end{tabular}
\caption{Comparison of different confidence intervals. The first column gives the level of the confidence interval. The other two columns show the proportion of data points inside the confidence interval.}
\label{citable}
\end{table}

\begin{table}[H]
\centering
\begin{tabular}{rrr}
  \hline
 & SAR & tSAR \\ 
  \hline
90$\%$ &0.1622 &0.7853 \\ 
95$\%$ &0.2566  &0.4265  \\ 
99$\%$ &0.0002 &0.8827  \\  
   \hline
\end{tabular}
\caption{Comparison of different confidence intervals. The first column gives the level of the confidence interval. The other two columns show the $p$-value of the likelihood ratio test.}
\label{pval}
\end{table}

\section{Outlook}

We proposed the tSAR model, an extension of the SAR model for $t$-distributed errors, which lead to notable improvements in the model fit in our application. The tSAR model showed improvement in the BIC value, its residuals behaved well and it provided more accurate confidence intervals. A natural question which arises is if we can extend the SAR model to other distributions than the \(t\)-distribution. Having a closer look at how we approached the tSAR model we can proceed in a similar way for other distributions. We consider the model
\begin{equation*}
\boldsymbol Y = X \boldsymbol\beta + \lambda W(\boldsymbol Y - X \boldsymbol\beta) + \boldsymbol\epsilon ,
\end{equation*}
where everything except \(\boldsymbol \epsilon\) is defined as in the SAR model (see Definition \ref{sardef}). 
We make the more general assumption for the error \(\boldsymbol \epsilon\) that it has expectation zero, a diagonal variance matrix \(\sigma^2 \Sigma_{\epsilon}\) and that \(\epsilon_i/(\sigma\sqrt{(\Sigma_{\epsilon})_{ii}})\) are identically and independent distributed with density \(\phi(\cdot |\boldsymbol\theta)\), where \(\phi(\cdot |\boldsymbol \theta)\) is the density of a distribution with zero mean, unit variance and parameter vector \(\boldsymbol\theta\). So one could allow for errors that follow for example a skew-$t$ distribution. Note that \(\boldsymbol\theta\) is empty for location-scale distributions (e.g. the normal distribution). We obtain the density of \(\boldsymbol Y\) as in Section \ref{paresttsar} using the density transformation rule as
\begin{equation*}
f_Y(\boldsymbol y) = |\det(\Sigma_{\epsilon}^{-\frac{1}{2}} (\Id_n - \lambda W))| \prod_{i=1}^n \phi\left( \left(\Sigma_{\epsilon}^{-\frac{1}{2}}(\Id_n - \lambda W) (\boldsymbol y - X \boldsymbol\beta)\right)_i | 0,1,\boldsymbol\theta\right).
\end{equation*}
The regression parameters \(\boldsymbol \beta\) could be estimated by the generalized least squares estimator and \(\sigma^2\) as in the SAR model. Then we can form the profile log-likelihood and estimate \(\lambda\) and \(\boldsymbol\theta\) by numerical optimization. Alternatively one could think about finding estimators of \(\boldsymbol\theta\) depending on \(\lambda\) such that the dimensionality of the profile log-likelihood can be reduced. It would be interesting to investigate this in more detail for various distributions.

\section*{Acknowledgment}

The first author acknowledges financial support by a research stipend of the Technical University of Munich. The second author is supported by the German Research Foundation through the TUM International Graduate School of Science and Engineering (IGSSE).
The  third  and  fourth  authors  are  supported  by  the  German  Research  Foundation  (DFG grants CZ 86/5-1 and CZ 86/4-1). Computations were performed on a Linux cluster supported by DFG grant INST 95/919-1 FUGG.

\bibliographystyle{plain}

\bibliography{MA_paper_bib}{}

\end{document}